\documentstyle[multicol,prb,aps,graphics,epsf]{revtex}

\begin{document}
\title{The transition from the adiabatic to the sudden limit in
core level photoemission:\\
A model study of a localized system}
\author{J.D. Lee and O. Gunnarsson}
\address{Max-Planck Institut f\"{u}r Festk\"{o}rperforschung, \\
Heisenbergstrasse 1, D-70569 Stuttgart, Germany}
\author{L. Hedin}
\address{Department of Theoretical Physics, University of Lund, \\
S\"{o}lvegatan 14 A, S-223 62 Lund, Sweden}
\date{\today}
\maketitle

\begin{abstract}
We consider core electron photoemission in a localized system, where
there is a charge transfer excitation. Examples are transition metal 
and rare earth compounds, chemisorption systems, and high 
$T_{c}$ compounds. The system is modelled by three electron levels, 
one core level and two outer levels. In the initital state the core 
level and one outer level is filled (a spinless two-electron problem). 
This model system is embedded in a solid state environment, and the 
implications of our model system results for solid state photoemission 
are disussed. When the core hole is created, the more localized outer 
level (d) is pulled below the less localized level (L).  The spectrum 
has a leading peak corresponding to a charge transfer between L
and d (''shake-down''), and a satellite corresponding to no charge
transfer.  The model has a Coulomb interaction between these levels 
and the continuum states into which the core electron is emitted. 
The model is simple enough to allow an exact numerical solution, 
and with a separable potential an analytic solution. Analytic results 
are also obtained in lowest order perturbation theory, and in the high 
energy limit of the semiclassical approximation. We calculate the 
ratio $r(\omega )$ between the weights of the satellite and the main 
peak as a function of the photon energy $\omega $. The transition 
from the adiabatic to the sudden limit is found to take place for  
quite small kinetic energies of the photoelectron. For such small
energies, the variation of the dipole matrix elements is substantial 
and described by the energy scale $\tilde{E}_{d}$. Without
the coupling to the photoelectron, the corresponding ratio 
$r_{0}(\omega )$ shows a smooth turn-on of the satellite intensity, 
due to the turn on of the dipole matrix element. The characteristic 
energy scales are $\tilde{E}_{d}$ and the satellite excitation energy 
$\delta E$. When the interaction potential with the continuum states 
is introduced a new energy scale $\tilde{E}_{s}=1/(2\tilde{R}_{s}^{2})$ 
enters, where $\tilde{R}_{s}$ is a length scale of the interaction 
(scattering) potential. At threshold there is typically a (weak) 
{\it constructive} interference between intrinsic and extrinsic 
contributions, and the ratio $r(\omega )/r_{0}(\omega )$ is larger 
than its limiting value for large $\omega $. The interference becomes 
small or weakly destructive for photoelectron energies of the order
$\tilde{E}_{s}$. For larger photoelectron energies $r(\omega )/r_{0}
(\omega )$ therefore typically has a weak undershoot. If this 
undershoot is neglected,  $r(\omega)/r_0(\omega)$ reaches its 
limiting value on 
the energy scale $\tilde E_s$ for the parameter range considered here.  
In a ''shake-up'' scenario, where the two outer levels do not cross 
as the core hole is created, we instead find that $r(\omega )
/r_{0}(\omega )$ is typically reduced for small $\omega $ by 
interference effects, as in the case of plasmon excitation. 
For the ''shake-down'' case, however, the results are very different 
from those for a simple metal, where plasmons dominate the picture. 
In particular, the adiabatic to sudden transition takes place at
much lower energies in the case of a localized excitation.
The reasons for the differences are briefly discussed.

\end{abstract}

\pacs{PACS numbers: 71.20.Be, 78.20.Bh, 79.60.-i}

\begin{multicols}{2}

\section{Introduction}

X-ray photoemission spectroscopy PES is a useful tool for studying the
electronic structure of solids. The theoretical description of PES is
however very complicated\cite{Hedinbook,Schaich} and almost all work has
been based on the so-called sudden approximation.\cite{sudden,Lars98}
The photoemission spectrum is then described by the electron spectral
function convoluted by a loss function, describing the transport of the 
emitted electron to the surface. The sudden approximation becomes exact 
in the limit when the kinetic energy of the emitted electron becomes 
infinite.\cite {sudden,Lars98} In this limit we can distinguish between 
intrinsic satellites, appearing in the electron spectral function, and 
extrinsic satellites, appearing in the loss function. For lower kinetic 
energy, this distinction is blurred due to interference effects, and 
the satellite weights are expected to be quite different as we approach 
the opposite limit, the adiabatic limit, of low kinetic 
energy.\cite{sudden} It is then interesting to ask at what kinetic 
energy the sudden approximation becomes accurate. This issue has been 
studied extensively for the case when the emitted electron couples to 
plasmons, and it has been found that the sudden approximation becomes 
valid only for very large ($\sim $ keV) kinetic 
energies.\cite{Lars98,plasmon,plasmonexp}

A semi-classical approach has been found to work exceedingly well for
the study of plasmon satellites.\cite{Lars98,plasmon} In such a picture,
one may take the emitted electron to move as a classical particle away 
from the region where the hole was created. The system then sees the 
potential from both the created electron and hole. Initially the 
electron potential cancels the hole potential, but as the electron 
moves away, the hole potential is gradually switched on. The 
switching-on of the hole potential may lead to the creation of 
excitations. If the kinetic energy of the electron is sufficiently 
large, we can consider the hole potential as being switched on
instantly, and the creation of excitations around the hole then reaches
a limiting value, the sudden limit. In the semi-classical picture, the
perturbation is turned on during a time $\tau =R_{0}/v$, where $v$ is
the photoelectron velocity and $R_{0}$ is the range of the interaction
(scattering) potential between the emitted electron and the excitations.
In this picture we also need to determine the relevant time scale 
$\tau _{{\rm max}}$ so that the sudden limit is reached if $\tau \ll 
\tau _{{\rm max}}$ or $v\gg R_{0}/\tau _{{\rm max}}$. Our analysis 
within the semi-classical framework shows that $1/\tau _{{\rm max}}$ 
is related to the energy $\delta E $ of the relevant excitation of 
the system and to the strength $\tilde{V}$ of the scattering potential.

We find a different characteristic energy scale $\tilde E_s= 1/(2 
\tilde R_s^2)$, where $\tilde R_s$ is a characteristic length scale 
of the scattering potential. On dimensional grounds one may argue that 
the adiabatic-sudden approximation takes place when the kinetic energy 
of the emitted electron is comparable to $\tilde E_s$. This would 
differ dramatically from the semi-classical approach, where the 
transition takes place for energy of the order $1/(\tilde E_s 
\tau_{{\rm max}}^2)$, i.e.  e.g., $(\delta E)^2/\tilde E_s$ or 
$\tilde V^2/\tilde E_s$. Alternatively, and again on dimensional 
grounds, one may argue that the sudden approximation becomes valid 
when the kinetic energy of the emitted electron is much larger than 
the energy $\delta E$ of the relevant excitations of the
system,\cite{Krause1974} in strong contrast to the two criteria above.
This latter criterion is however not true in general.\cite{Shirley1987}

\begin{figure}
\vspace*{4cm}
\includegraphics{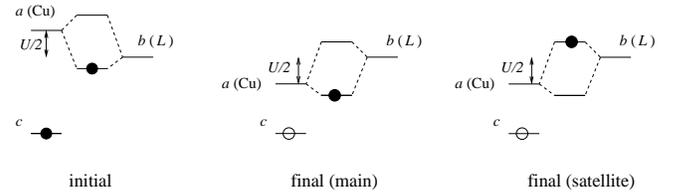}
\caption{Schematic view of the Cu $3s$ charge transfer photoemission.
Here $a
$ is the Cu $3d$ level and $b$ is a ligand (L) valence state
for the symmetric case.
}
\label{fig:schematic}
\end{figure}

For many systems with strong correlations, the core level spectrum can
be understood in a charge transfer scenario.\cite{Kotani} This is
illustrated in Fig. 1 for a Cu compound, e.g., a Cu halide. In the 
ground-state, Cu has
essentially the configuration d$^{9}$ and all the ligand orbitals are
filled. In the presence of a Cu core hole, it becomes energetically
favorable to transfer an electron from a ligand to the d-shell,
obtaining a $%
d^{10}$ configuration on the Cu atom with the core hole. Due to the
hybridization between the d$^{9}$ and $d^{10}$ configurations, the
states
are actually mixtures of the two configurations, as indicated in Fig.
\ref
{fig:schematic}. In the photoemission process there is a nonzero
probability
that the outer electron will not stay on the ligand, but is transferred
to
the lower energy $d$-like state. This ``shake-down''  process
corresponds to
the leading peak in the spectrum, while the process where the outer
electron
stays on the ligand corresponds to the satellite. This kind of model has
been applied to rare earth compounds,\cite{Kotani,Gunnar83}
chemisorption
systems,\cite{Schonhammer,Fuggle} transition metal compounds,\cite
{Larsson,Wallbank,Laan,Boer,Fujimori} and High $T_{c}$ compounds.\cite
{Veenendaal}

Our simple model allows an accurate numerical calculation of the
photocurrent either by integrating the time-dependent Schr\"{o}dinger
equation
or by directly inverting a resolvent operator (QM). We also derive
analytic
results with a separable potential. These results are compared with the
semi-classical theory (SC), and with first order perturbation theory
(PT).
In both these cases we have analytic results, which is very useful for
understanding the physics of the problem.

The impurity model discussed here differs in certain important aspects
from a real solid.  To start with, for a solid we never reach the 
limit of a pure intrinsic spectrum since when the cross section for 
extrinsic scattering goes to zero, the range from which the 
photoelectrons come goes to infinity. In our impurity model, on the 
other hand, the extrinsic scattering approaches zero at high kinetic 
energies. Secondly, for a solid, we discuss excitations in
the continuum and not as here discrete energy levels.

For the coupling to plasmons, the adiabatic-sudden transition takes
place at large kinetic energies where the SC approximation is very 
accurate.\cite{Lars98,plasmon} The relevant length scale is given 
by the plasmon wave length $\lambda =2\pi /q$ and the relevant 
time by the inverse plasmon frequency $\omega _{q}$. Large 
interference effects are then connected with a large phase velocity 
$\omega _{q}/q$, as discussed e.g. by Inglesfield.\cite{Inglesfield} 
Since long wave-length plasmons play an important role these
large interference effects for small $q$ delay the approach to the
sudden limit, which only is reached at very high kinetic energies 
($\sim $ keV).

For the localized excitations studied here the relevant length scale is
much
shorter. The SC theory then predicts that the transition takes place at
correspondingly smaller kinetic energies. This is indeed what we find
from
the exact solution of our model. This has two consequences. Firstly, the
SC
treatment itself is not valid at such small energies, and we have to
rely on
QM treatments. Actually, although the SC treatment correctly predicts a
small transition energy, we find it predicts qualitatively wrong
dependencies on the relevant parameters. Secondly, the smaller energy
scale
means that the energy variation of the dipole matrix elements becomes
very
important. The dipole matrix element grows rapidly on an energy scale $%
\tilde E_d$, which can become very important for the adiabatic-sudden
transition.

We study the ratio $r(\omega )$ between the weights of the satellite 
and the main peak as a function of the photon energy $\omega $ for 
the emission from a $3s$ level. First we consider the case when the 
scattering potential between the electron and the target is neglected. 
We find that the corresponding ratio $r_{0}(\omega )$ strongly depends 
on the ratio between the excitation energy $\delta E$ and $\tilde{E}_{d}$. 
If $\delta E/\tilde{E}_{d}\ll 1$, $r_{0}(\omega )$ approaches its 
limiting value from below, while it has an overshoot if 
$\delta E/\tilde{E}_{d}\gg 1$. In both case the limit value is reached 
for photoelectron energies of the order of a few times $\delta E$.

We then study the effects of the scattering potential by focusing on $%
r(\omega )/r_{0}(\omega )$. For small energies there is typically an
overshoot due to {\it constructive} interference in the ''shake-down''
case, contrary to the shake-up case where, like for plasmons, $r(\omega
)/r_{0}(\omega )$ is reduced by interaction effects. This happens on the
energy scale $\tilde{E}_{s}$. If the scattering potential is very
strong, this overshoot may extend to several times $\tilde{E}_{s}$.
Depending on the parameters there may be an undershoot for higher
energies, which can extend up to quite high energies. The undershoot
is, however, rather small for the parameters we consider here,
and should therefore not be very important unless we want to calculate
the spectrum with a high accuracy. For Cu compounds and emission
from the $3s$ core level, we find that $\tilde E_d$ and $\tilde E_s$ are
comparable, and the relevant energy for $r(\omega)/r_0(\omega)$
is then given by $\tilde E\sim \tilde E_s\sim \tilde E_d$.

We present our model in Sec. II and calculate various matrix elements in
Sec. III. The sudden approximation is described in Sec. IV and exact
numerical methods are given in Sec. V. The perturbational and the
semi-classical treatments are presented in Sec. VI. In Sec. VII we study
the
condition for the adiabatic-sudden transition qualitatively, using
simple
analytic matrix elements and within the framework of the semi-classical
theory. The results are discussed in Sec. VIII.

\section{Model}

We consider a Hamiltonian ${\cal H}_{0}$ describing a model with a core
level $c$ and two valence levels $a$ and $b$,
\begin{eqnarray}\label{eq:1}
{\cal H}_{0} &=&\epsilon _{a}n_{a}+\epsilon _{b}n_{b}+\epsilon
_{c}n_{c}+U_{a}n_{c}n_{a}+U_{b}n_{c}n_{b}  \nonumber \\
&+&t(c_{a}^{\dagger }c_{b}+c_{b}^{\dagger }c_{a}).
\end{eqnarray}
The first two terms give the bare energies of the levels $a$ and $b$,
and
the last term the hybridization between them. The remaining terms
involve
the occupation number $n_{c}$ of the core level $c$. In photoemission
the
core level is filled in the initial state, and empty in the final, and
$n_{c}
$ only enters as a constant. It is trivial to diagonalize ${\cal
H}_{0}$,
and one obtains two dressed energies $E_{a}(n_{c})$ and $E_{b}(n_{c})$
for
the levels $a$ and $b.$ In a Cu compound, for instance, $c$ may
represent
the Cu $3s$ core level, $a$ the Cu $3d$ valence level and $b$ a ligand
state. This is schematically illustrated in Fig. \ref{fig:schematic}. In
our
calculations we almost always treat the case when $E_{a}(1)>E_{b}(1)$,
and $%
E_{b}(0)>E_{a}(0)$. The meaning of the levels $a$ and $b$ for different
types of systems with localized excitations is indicated in Table \ref
{general-CT}. The full Hamiltonian also has a one-electron part for
continuum states,
\begin{equation}
T=\sum_{{\bf k}}\epsilon _{{\bf k}}n_{{\bf k}},  \label{eq:2}
\end{equation}
with the energies $\epsilon _{{\bf k}}=k^{2}/2$, and wavefunctions
$\psi
_{{\bf k}}$ obtained from a one-electron potential corresponding to
$n_{c}=0$%
. The $n_{{\bf k}}$ are occupation numbers $n_{{\bf k}}=c_{{\bf
k}}^{\dagger
}c_{{\bf k}}$. We use atomic units with $e=m=\hbar =1,$ and thus, e.g.
energies are in hartrees ($27.2$ eV). The perturbation causing
photoemission is
\begin{equation}
\Delta =\sum_{{\bf k}c}(M_{{\bf k}}c_{{\bf k}}^{\dagger }c_{c}+h.c.),
\label{eq:3}
\end{equation}
where $M_{{\bf k}}$ is an optical transition matrix element. We take the
photoelectron interaction as
\begin{equation}
V=\sum_{{\bf k}{\bf k}^{^{\prime }}}[n_{a}V_{{\bf k}{\bf k}^{^{\prime
}}}^{(a)}+n_{b}V_{{\bf k}{\bf k}^{^{\prime }}}^{(b)}-V_{{\bf k}{\bf k}%
^{^{\prime }}}^{(c)}]c_{{\bf k}}^{\dagger }c_{{\bf k}^{^{\prime }}}.
\label{eq:5}
\end{equation}
Here $V_{{\bf k}{\bf k}^{^{\prime }}}^{(\nu )}$ is a matrix element of
the
Coulomb potential $V^{(\nu )}({\bf r})$ from the charge density $\rho
_{\nu
}({\bf r})$ of the orbital $\nu $,
\[
V_{{\bf k}{\bf k}^{^{\prime }}}^{(\nu )}=\int \psi _{{\bf k}}^{\ast
}\left(
{\bf r}\right) V^{(\nu )}({\bf r})\psi _{{\bf k}^{\prime }}\left( {\bf
r}%
\right) d{\bf r,\;}V^{(\nu )}({\bf r})=\int {\frac{\rho _{\nu }({\bf r}%
^{\prime })}{|{\bf r}-{\bf r}^{\prime }|}}d{\bf r}^{\prime }.
\]
It is the potential $V$ which determines the transition from the
adiabatic to the sudden limit, with $V=0$ we are in the sudden limit.

The total Hamiltonian is given by
\begin{equation}
{\cal H}={\cal H}_{0}+T+V+\Delta .  \label{eq:6}
\end{equation}
This Hamiltonian has two conserved quantities,
\begin{equation}
n_{c}+\sum_{{\bf k}}n_{{\bf k}}=1\ \ \ {\rm and}\ \ \ n_{a}+n_{b}=1.
\label{eq:7}
\end{equation}
For simplicity we take the core electron and the $d$ electron potentials
as equal, $V^{c}=V^{a}$, and use the relation $n_{a}+n_{b}=1$ to obtain
\begin{equation}
V=n_{b}\sum_{{\bf k}{\bf k}^{^{\prime }}}V_{{\bf k}{\bf k}^{^{\prime
}}}c_{ {\bf k}}^{\dagger }c_{{\bf k}^{^{\prime }}}  \label{eq:8}
\end{equation}
where
\begin{eqnarray*}
V_{{\bf kk}^{^{\prime }}} &\equiv &\int \psi _{{\bf k}}^{\ast }\left(
{\bf r} \right) V_{sc}({\bf r})\psi _{{\bf k}^{\prime }}
\left( {\bf r}\right) d{\bf r,\;} \\
V_{sc}({\bf r}) &\equiv &V^{(b)}({\bf r})-V^{(a)}({\bf r})=\int d{\bf
r} ^{\prime }{\frac{1}{|{\bf r}-{\bf r}^{\prime }|}}[\rho _{b}({\bf
r}^{\prime })-\rho _{a}({\bf r}^{\prime })].
\end{eqnarray*}
The (scattering) potential $V_{sc}({\bf r})$ describes the change in the
potential acting on the emitted electron when the electron in the target
hops from level $a$ to $b$. 

Dropping a constant we can write ${\cal H}_{0}$ as
\begin{eqnarray}\label{eq:9}
{\cal H}_{0} &=&\epsilon _{c}n_{c}+(\epsilon _{a}+Un_{c})n_{a}+\epsilon
_{b}n_{b}  \nonumber   \\
&+&t(c_{a}^{\dagger }c_{b}+c_{b}^{\dagger }c_{a}).
\end{eqnarray}
where the Coulomb integral $U$ is given by
\begin{eqnarray}\label{eq:m1}
U &\equiv &U_{a}-U_{b}  \nonumber   \\
&=&\int d{\bf r}d{\bf r}^{\prime }\rho _{c}({\bf r}){\frac{1}{|{\bf
r}-{\bf r%
}^{\prime }|}}[\rho _{a}({\bf r}^{\prime })-\rho _{b}({\bf r}^{\prime
})].
\end{eqnarray}
Since the core level is very localized in space this leads to
\begin{equation}
U=-V_{sc}(0).  \label{eq:m3}
\end{equation}
For the different types of systems in Table \ref{general-CT}, $a$ refers
to
a localized level and $b$ refers to a more extended level. For instance,
for
a copper dihalide compound, $a$ refers to a Cu $3d$ orbital and $b$ to a
combination of orbitals on the ligand sites. For simplicity, we
approximate
the six ligand orbitals by a spherical shell with the radius
$R_{0}$,\cite
{Kaupp} where $R_{0}$ is the average Cu-ligand separation. The potential
from the Cu $3d$ orbital $V_{3d}(r)$ can be considered as purely
Coulombic
at $r=R_{0}$. The charge from the spherical shell gives a constant
potential
inside the radius $R_{0}$, and we have

\begin{equation}  \label{eq:m4}
V_{sc}(r)=\left\{
\begin{array}{ll}
(-V_{3d}(r)+\frac{1}{R_{0}})/\varepsilon & r<R_{0}; \\
0 & r>R_{0}.
\end{array}
\right.
\end{equation}
Here $\varepsilon $ is a constant chosen to make $U=-V_{sc}(0)$, which
may be thought of as being due to screening by the surrounding. Since 
$V_{3d}(0)\gg 1/R_{0}$, $\varepsilon $ varies only weakly with $R_{0}$
and is approximately given by $\varepsilon \simeq V_{3d}(0)/U$.

\section{Matrix elements}

\label{sec:matrix}

To estimate the matrix elements $M_{{\bf k}}$ and $V_{{\bf k}{\bf k}%
^{^{\prime }}}$ we must approximate the photoelectron wavefunctions
$\psi _{{\bf k}}\left( {\bf r}\right) $. These wavefunctions are 
calculated from the potential of a neutral atom, which further is 
shifted to make the potential zero outside a muffin-tin radius 
$r_{mt}$. The states are then described by spherical Bessel functions 
outside $r_{mt}$, which are matched to a solution
of the atomic potential inside $r_{mt}$. For the energy $k^{2}/2$ we
obtain
the partial wave
\begin{equation}
R_{lk}(r)=\cases {a_{lk}\psi_{lk}(r) & $r<r_{mt}$;\cr
\sqrt{\frac{2}{R}}k[\cos\eta_{lk}j_l(kr)-\sin\eta_{lk}n_l(kr)] &
$r>r_{mt}$,\cr}  \label{eq:ma1}
\end{equation}
where $\psi _{lk}(r)$ is the solution of the radial Schr\"{o}dinger
equation
for the atomic potential inside the muffin-tin radius, $a_{lk}$ is a
matching coefficient and $\eta _{lk}$ a phase shift. The normalization
is
given by
\begin{equation}
\int_{0}^{R}drr^{2}R_{lk}^{2}(r)=1,  \label{eq:ma1a}
\end{equation}
where $R$ is the radius of a large sphere to which the continuum states
are
normalized. The factor $(2/R)^{1/2}k$ is due to the normalization and
the
asymptotic behavior of $j_{l}(x)$ for large $x$.

\begin{figure}
\vspace*{10.5cm}
\includegraphics{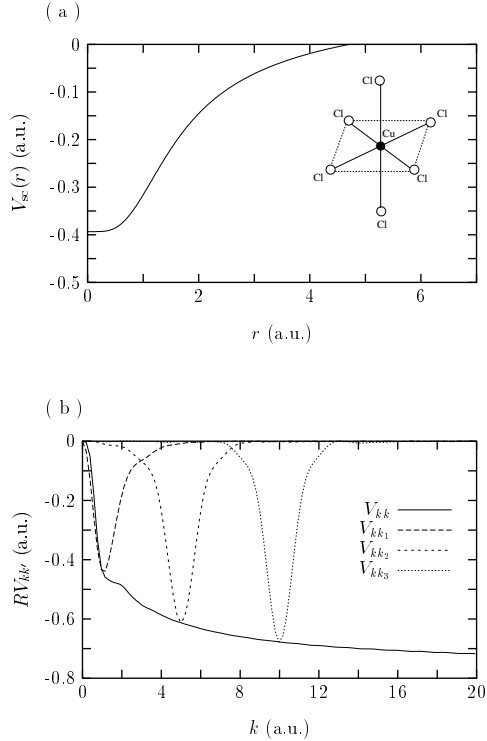}
\caption{(a) The photoelectron scattering potential $V_{sc}(r)$ given by
Eq.
(\ref{eq:m4}) with respect to $r$ for CuCl$_2$ ($R_0=4.71$ a.u. and
$\protect%
\varepsilon=1.96$). In the inset, we give the atomic configuration of
Cu-Cl
octahedral (nearly octahedral) cluster in CuCl$_2$. (b) The diagonal and
off-diagonal matrix elements of the scattering potential multiplied by
$R$.
In the figure, $k_1=1$ a.u., $k_2=5$ a.u., and $k_3=10$ a.u. are taken.
}
\label{fig:potential}
\end{figure}

\begin{figure}
\vspace*{10.5cm}
\includegraphics{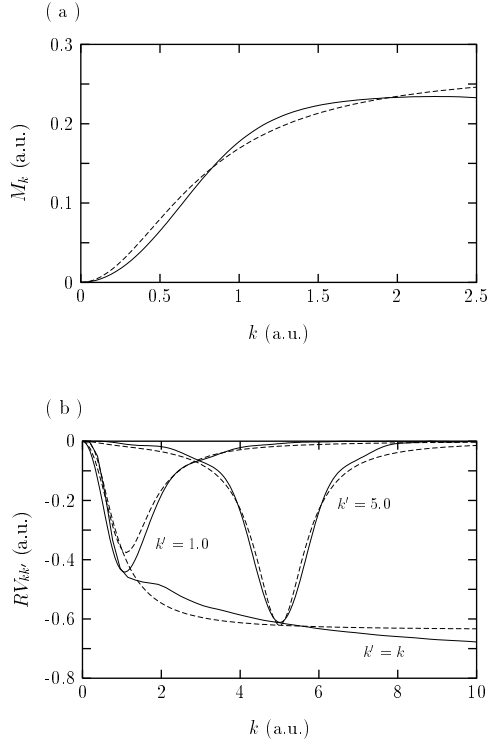}
\caption{(a) The dipole matrix element $M_{k}$ as a function of $k$. The
exact result (solid line) is obtained from Eq. (\ref{eq:ma1b}) and the
simplified model by Eq. (\ref{eq:ma4}) is also shown (dashed line). The
appropriate parameter is $\tilde R_d=1.30$ a.u. (b) The matrix elements
$%
V_{kk^{^{\prime}}}$ of the scattering potential are given for $%
k=k^{^{\prime}}$, $k^{^{\prime}}=1$ a.u., and $k^{^{\prime}}=5$ a.u..
The
solid line is from the exact calculation for the model of CuCl$_2$ and
the
dashed line is based on the simplified model (Eq.(\ref{eq:ma6})). The
parameters are $\tilde{V}=-0.36$ a.u., $\tilde R_s=1.77$ a.u. and
$\tilde R _{sd}=1.31$ a.u.. }
\label{fig:dipole}
\end{figure}

Slater's rules\cite{Slater} are used to generate the orbitals and charge
densities, from which the potential $V_{3d}(r)$ is calculated. This
gives
the scattering potential $V_{sc}(r)$, which is shown in Fig. \ref
{fig:potential}a. We consider photoemission from a Cu $3s$ hole. Due to
the
dipole selection rules, the core electron is then emitted into a
continuum
state of $p$-symmetry. The matrix elements $V_{{k}{k}^{^{\prime }}}$ of
the
scattering potential $V_{sc}(r)$ are shown in Fig. \ref{fig:potential}b,

\begin{equation}
V_{kk^{^{\prime }}}=\int drr^{2}R_{k}(r)V_{sc}(r)R_{k^{^{\prime }}}(r),
\label{eq:ma0}
\end{equation}
where the muffin-tin radius $r_{mt}$ is taken as the ionic radius of Cu,
$%
r_{mt}=2.6$ a.u., and we have dropped the $l$ index, since we always
consider $l=1.$ The dipole matrix element $M_{k}$ is given by
\begin{equation}
M_{k}\sim a_{k}(\epsilon _{k}-\epsilon _{c})\int drr^{2}\psi
_{3s}(r)r\psi
_{k}(r).  \label{eq:ma1b}
\end{equation}
We assume that the core level is deep, and $\left| \epsilon_{c}\right|$
much larger than the energy difference between the ligand and copper
levels.  We can then take the factor $\epsilon_{k}-\epsilon_{c}$ 
as a constant, which drops out since we always consider relative 
intensities. The result for $M_{k}$ is shown as the solid line in 
Fig. \ref{fig:dipole}a.  These dipole and scattering potential
matrix elements are used in the following numerical calculations.
Extensive calculations of dipole matrix elements for many systems 
were performed by Yeh and Lindau.\cite{crossection} 

To interpret the results, it is useful to also perform analytical
calculations. For this purpose we need models of the matrix elements.
Below
we consider the limits of low and high kinetic energies of the emitted
electron. In the limit of low kinetic energies, we replace the spherical
Bessel function by its expansion for small arguments
\begin{equation}  \label{eq:ma2}
j_l(x)=\frac{1}{(2l+1)!!}x^l,\mbox{ }\mbox{ } n_l(x)=-(2l-1)!!
\frac{1}{ x^{l+1}}
\end{equation}
and the solution $\psi_{lk}(r)$ by its zero energy limit $\psi_{l0}(r)$.
This leads to
\begin{eqnarray}  \label{eq:ma3}
\tan\eta_{lk}=\frac{l-\xi}{l+1+\xi}\frac{(kr_{mt})^{2l+1}}
{(2l-1)!!(2l+1)!!} \sim\eta_{lk},  \nonumber \\
a_{lk}=\sqrt{\frac{2}{R}}k\left[\frac{2l+1}{l+1+\xi}
\frac{(kr_{mt})^l}{ (2l+1)!!}\right]\frac{1}{\psi_{l0}(r_{mt})}.
\end{eqnarray}

Due to the matching, the coefficient $a_{lk}$ contains the ratio $%
\xi=r_{mt}\psi^{^{\prime}}_{l0}(r_{mt})/\psi_{l0}(r_{mt})$. The value of
the
coefficient therefore depends in an interesting way on the wave function
$%
\psi_{l0}$ and its derivative, and if $l+1+\xi$ is close to zero
$a_{lk}$
blows up. Then the matrix elements of the scattering potential also blow
up
and we may expect strong deviations from the sudden approximation. In
such a
case the dipole matrix element $M_{lk}$ also becomes large, i.e., there
is a
resonance ($\eta_{lk}=\frac{\pi}{2}$) in the photoemission cross
section.

 From Eqs.(\ref{eq:ma1}) and (\ref{eq:ma1b}) it follows that in the limit
of
a small $k$ the dipole matrix element is proportional to $a_{k}\sim
k^{l+1}$%
. The main peak and the satellites in the photoemission spectrum
correspond
to different kinetic energies and therefore have dipole matrix elements
with
different $k$-values. This is important at low energies, while for large
energies the variation in $M_{lk}$ is generally small over a range of
the
energy difference between the main peak and the satellite, and for the
ratio
of the peaks, the dipole matrix elements should then not play a role.
For
simplicity, we assume that the dipole matrix elements become independent
of $%
k$ for $\tilde{R}_{d}k\gg 1$, where $\tilde{R}_{d}$ is some typical
length
scale of the system. For our case ($l=1$) we use the model (note that
any
constant factor in $M_{k}$ drops out in our final expressions)
\begin{equation}
M_{k}={\frac{(\tilde{R}_{d}k)^{2}}{1+(\tilde{R}_{d}k)^{2}}}\equiv
{\frac{%
\epsilon _{k}/\tilde{E}_{d}}{1+\epsilon _{k}/\tilde{E}_{d}}},
\label{eq:ma4}
\end{equation}
where $\tilde{E}_{d}=1/(2\tilde{R}_{d}^{2})$. Fig. \ref{fig:dipole}a
compares this model with the full calculation for a $3s$ orbital. We
obtain $%
\tilde{R}_{d}=1.3$ a.u.. For a $1s$ or $2s$ orbital the length scale is
smaller and $\tilde{R}_{d}\sim 1/2$ a.u.. While we consider $l=1$ the
behaviour for other $l$ is primarily modified for small energies.

We next consider the matrix elements $V_{kk^{^{\prime}}}$. For small
values
of $k$ and $k^{^{\prime}}$ and for $l=1$
\begin{eqnarray}  \label{eq:ma5}
V_{kk^{^{\prime}}}&=&\frac{2}{R}\left[\frac{r_{mt}}{\psi_{l0}(r_{mt})}\right]
^2
\frac{(kk^{^{\prime}})^2}{(\xi+2)^2}\int_0^{r_{mt}}\psi_{l0}^2(r)V_{sc}
(r) r^2dr  \nonumber \\
&+&\frac{2}{9R}(kk^{^{\prime}})^2\int_{r_{mt}}^{R_0}
\left[r+\left(\frac{
1-\xi}{2+\xi}\right)\frac{r_{mt}^3}{r^2}\right]^2 V_{sc}(r)r^2dr.
\nonumber
\\
\end{eqnarray}
For small values of $k$ and $k^{^{\prime}}$ it then follows that $%
V_{kk^{^{\prime}}}\sim (kk^{^{\prime}})^2$. For large values of $k$ and
$%
k^{^{\prime}}$ the matrix elements become very small due to destructive
interference between the two wave-functions unless $k\approx
k^{^{\prime}}$.
If $k=k^{^{\prime}}$ the matrix elements $V_{kk}$ approach a constant.
These
features are contained in the model
\begin{equation}  \label{eq:ma6}
V_{kk^{^{\prime}}}={\frac{\tilde V \tilde R_s}{R}}{\frac{ (\tilde R%
_s^2kk^{^{\prime}})^2 }{ \lbrack 1+(\tilde R_s k)^2\rbrack \lbrack
1+(\tilde %
R_s k^{^{\prime}})^2\rbrack \lbrack 1+\tilde
R_{sd}^2(k-k^{^{\prime}})^2%
\rbrack}},
\end{equation}
where $\tilde R_s$ and $\tilde R_{sd}$ are appropriate length scales and
$%
\tilde V$ has the energy dimension. We recall that $\tilde V$ contains
information about the coefficient $a_{lk}$ defined in Eq.(\ref{eq:ma3})
and
therefore about the atomic potential and that $\tilde V$ may become
particularly large close to a resonance. The dipole matrix element $M_k$
and
potential matrix $V_{kk^{^{\prime}}}$ in our simplified model Eqs. (\ref
{eq:ma4}) and (\ref{eq:ma6}) are given in Fig. \ref{fig:dipole} as
compared
with the exact results.

For large values of $k$ and $k^{^{\prime }}$, the expression Eq.~(\ref
{eq:ma6}) simplifies to
\begin{equation}
V_{kk^{^{\prime
}}}={\frac{\tilde{V}\tilde{R}_{s}}{R}}{\frac{1}{1+\tilde{R}%
_{sd}^{2}(k-k^{^{\prime }})^{2}}}.  \label{eq:ma6c}
\end{equation}
An expression of this type can also be derived by assuming that the wave
functions $R_{lk}(r)$ can be approximated by spherical Bessel functions
in
all of space, and by assuming some shape of $V_{sc}(r)$, e.g., a linear
dependence on $r$

\begin{equation}
V_{sc}(r)=V_{sc}(0)(1-{\frac{r}{R_{0}}}).  \label{eq:ma6a}
\end{equation}
For large values of $k$ and $k^{^{\prime }}$ we then obtain
\begin{equation}
V_{kk^{^{\prime }}}={\frac{V_{sc}(0)R_{0}}{R}}{\frac{1-{\rm cos}%
[(k-k^{^{\prime }})R_{0}]}{[(k-k^{^{\prime }})R_{0}]^{2}}}.
\label{eq:ma6aa}
\end{equation}
For this model we relate $\tilde{R}_{s}=\tilde{R}_{sd}=R_{0}/3$ to the
range
$R_{0}$ of the potential and $\tilde{V}=3V_{sc}(0)/2$. Using this
identification in Eq. (\ref{eq:ma6c}), leads to the correct average
value of $V_{kk}$ and to the correct width in $k-k^{^{\prime }}$ of
$V_{kk^{^{\prime
}}}$. The simple form (\ref{eq:ma6c}), however, neglects the effects of
the oscillations of the cos-function in Eq. (\ref{eq:ma6aa}) for large
values of
$(k-k^{^{\prime }})R_{0}$, and it therefore gives a worse representation
of the linear potential (\ref{eq:ma6a}) than the form (\ref{eq:ma6}) gives
for the more realistic scattering potential (\ref{eq:m4}). Later we will
find that it is a reasonable approximation to put $\tilde{R}_{d}$,
$\tilde{R}_{sd}
$ and $\tilde{R}_{s}$ equal to the same value $\tilde{R}$, and introduce
the corresponding energy $\tilde{E}=1/(2\tilde{R}^{2})$.

\section{Sudden approximation}

\label{sec:sudden}

We first discuss the photoemission in the sudden limit, i.e., we neglect
the scattering potential between the emitted electron and the target
($V\equiv 0$). The initial state $|\Psi _{0}\rangle$ is the ground 
state of ${\cal H}_{0} $ with $n_{c}=1$ and given by
\begin{equation}  \label{eq:s1}
|\Psi_0\rangle=-\sin\theta |\psi_c\rangle|\psi_a\rangle +\cos\theta
|\psi_c\rangle|\psi_b\rangle,
\end{equation}
where
\begin{equation}
\tan 2\theta =2t/(\epsilon_{a}+U-\epsilon_{b})  \label{eq:s2}
\end{equation}
and the corresponding ground state energy is
\begin{equation}  \label{eq:s3}
E_0=\epsilon_c+{\frac{1}{2}}(\epsilon_a+U+\epsilon_b)-
\frac{1}{2}\sqrt{%
(\epsilon _{a}+U-\epsilon _{b})^{2}+4t^{2}}.
\end{equation}
The final states of the target are given by the two eigenstates of
${\cal H} _0$ with $n_c=0$
\begin{eqnarray}  \label{eq:s4}
|\psi _{1}\rangle &=&\cos \varphi |\psi _{a}\rangle -\sin \varphi |\psi
_{b}\rangle ,  \nonumber \\
|\psi _{2}\rangle &=&\sin \varphi |\psi _{a}\rangle +\cos \varphi |\psi
_{b}\rangle ,
\end{eqnarray}
with
\begin{equation}
\tan 2\varphi =2t/(\epsilon _{b}-\epsilon _{a})  \label{eq:s5}
\end{equation}
and the corresponding energy eigenvalues $E_{1}$ and $E_{2}$ are
\begin{equation}
E_{ {1  \atop 2} }={\frac{1}{2}}(\epsilon _{a}+\epsilon _{b})\mp 
\delta E/2, \label{eq:s6} \\
\end{equation}
with
\begin{equation}
\delta E=\sqrt{(\epsilon _{a}-\epsilon _{b})^{2}+4t^{2}}  \label{eq:s7}
\end{equation}
being the optical excitation energy of the system.

The photocurrent $J_{k}^{s}(\omega )$ ($s=1,2$) is given
by,\cite{Hedinbook}
\begin{equation}
J_{k}^{s}(\omega )=|\langle \Psi _{f}^{sk}|\Delta |\Psi _{0}\rangle
|^{2}\delta (\omega -\epsilon _{k}+E_{0}-E_{s}),  \label{eq:s8}
\end{equation}
where $|\Psi _{f}^{sk}\rangle $ is a final state. According to the
sudden
approximation, it can be written as the final target state multiplied by
the
photoelectron state, $|\Psi _{f}^{sk}\rangle =|\psi _{s}\rangle |\psi
_{k}\rangle $. This gives
\begin{equation}
\langle \Psi _{f}^{sk}|\Delta |\Psi _{0}\rangle =M_{k}w_{s}\equiv
m_{sk},\mbox{ }\mbox{ }w_{s}=\left\{ \begin{array}{c}
-\sin \left( \varphi +\theta \right) ,\;s=1 \\
\cos \left( \varphi +\theta \right) ,\;\;s=2
\end{array}
\right.   \label{eq:s9}
\end{equation}
$J_{{\bf k}}^{1}(\omega )$ gives the main line (corresponding to the
quasi
particle line in metal) and $J_{{\bf k}}^{2}(\omega )$ the satellite
line.
The schematic picture of the initial and final state for this system is
given in Fig. \ref{fig:schematic}. Summing the kinetic energy
distribution
of the photoelectron, we obtain the absorption spectra $J_{s}(\omega )$,

\begin{eqnarray} \label{eq:s10}
J_{s}(\omega ) &=&\sum_{k}J_{k}^{s}(\omega )\propto \frac{1}{k_{s}}%
|M_{k_{s}}w_{s}|^{2}  \nonumber  \\
k_{s} &=&\sqrt{2(\omega +E_{0}-E_{s})},
\end{eqnarray}
where the threshold energies for $J_{1}(\omega )$ and $J_{2}(\omega )$
are
given by $E_{1}-E_{0}$ and $E_{2}-E_{0}(\equiv \omega _{th})$,
respectively.
We can thus also write $k_{1}=\sqrt{2(\omega +\delta E-\omega _{th})}$
and $%
k_{2}=\sqrt{2(\omega -\omega _{th})}$. The factor $1/k$ comes from the
$k$%
-summation over a $\delta $-function in energy. For convenience we
introduce
the quantity $\tilde{\omega}=\omega -\omega _{th}$, and thus $\epsilon
_{k_{2}}=\tilde{\omega}$.

\begin{figure}[bt]
\unitlength1cm
\begin{minipage}[t]{3.125in}
\centerline{\rotatebox{-90}{\epsfxsize=2.5in
\epsffile{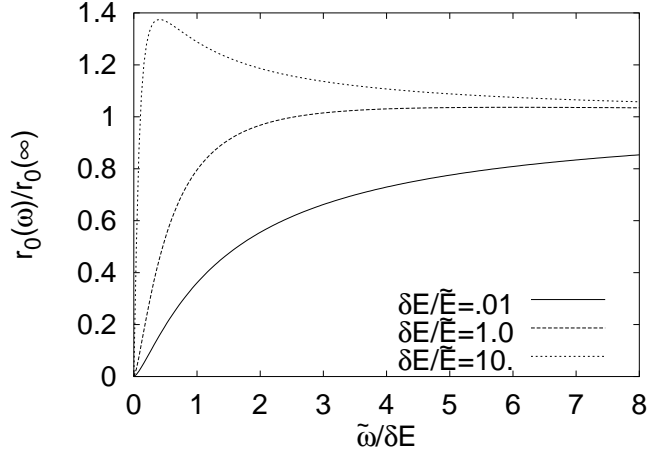}}}
\caption[]{\label{fig:sudden}The ratio $r_0(\omega)$ of the
satellite to the main peak  in Eq.(\ref{eq:s12}) divided
by the result for an infinite photon energy ($r_0(\infty)=r_{00}$). 
Three values of the excitation energy $\delta E$ are considered.}
\end{minipage}
\hfill
\end{figure}

In the sudden approximation the kinetic energy of the emitted electron
is large, and we can take $k_{1}=k_{2}$. The ratio $r_{00}$ of the
satellite to
the main peak intensity then is
\begin{equation}
r_{00}=\lim_{\omega \rightarrow \infty }\frac{J_{2}(\omega
)}{J_{1}(\omega )}%
=\cot ^{2}(\varphi +\theta ).  \label{eq:s11}
\end{equation}
Taking into account the energy dependence of the dipole matrix element
according to model Eq. (\ref{eq:ma4}) as well as the factor $1/k$, we
obtain
\begin{eqnarray}\label{eq:s12}
r_{0}(\omega ) &=&r_{00}\left[
{\frac{\tilde{\omega}}{\tilde{\omega}+\delta E%
}}\right] ^{\frac{3}{2}}  \nonumber  \\
&&\times \left[ {\frac{1+(\tilde{\omega}+\delta
E)/\tilde{E}_{d}}{1+\tilde{%
\omega}/\tilde{E}_{d}}}\right] ^{2}\Theta (\tilde{\omega}).
\end{eqnarray}
We now require that the ratio $r_{0}(\omega )$ should reach a fraction
$%
\gamma $ ($\gamma \approx 1$) of its limiting value $r_{0}(\infty )$ for
$%
\omega =\omega _{\gamma }$. This gives
\begin{equation}
{\frac{\omega _{\gamma }-\omega _{th}}{\delta E}}\approx \cases{{3\over
2}{1\over 1-\gamma}, & if $\delta E \ll \tilde{E}_d$;\cr {\gamma^{2/ 3}
(\tilde{E}_d/\delta E)^{4/3}}, & if $\delta E \gg \tilde{E}_d$.\cr}
\label{eq:s13}
\end{equation}
This criterion refers to the energy where $r_{0}(\omega )$ reaches a
fraction $\gamma $ in its rising part, and it does not consider that
there
is a large overshoot for $\delta E/\tilde{E}_{d}\gg 1$. In this case we
can
instead require that $r_{0}(\omega )$ is smaller than $\gamma \approx 1$
in
its descending part. This gives the condition
\begin{equation}
{\frac{\omega _{\gamma }-\omega _{th}}{\delta E}}\approx
{\frac{1}{\gamma
^{2}-1}}\hskip0.5cm\delta E/\tilde{E}_{d}\gg 1,\gamma >1.
\label{eq:s13a}
\end{equation}
In Fig. \ref{fig:sudden} we show results for $r_{0}(\omega )$ over a
large range of values for $\delta E/\tilde{E}$. The figure illustrates 
that the dipole matrix element effect alone makes the sudden 
approximation invalid for small kinetic energies. It is interesting 
that for somewhat larger photon energies $r_{0}$ overshoots. The 
reason is that the matrix elements $M_{k}$ saturate for $\epsilon
 _{k}\gg \tilde{E}_{d}$, while the factor $1/k$ in Eq.(\ref{eq:s10}) 
favors the satellite. For $\delta E/\tilde{ E}_{d}=1$, the result is 
rather close to the sudden limit for $\tilde{\omega}
/\delta E\sim 1$. Finally, for $\delta E/\tilde{E}_{d}\gg 1$, there is a
substantial overshoot.

As discussed in the introduction, we would like to study how the
adiabatic
to sudden transition depends on certain factors, like the range $R_{0}$
of
the potential and the energy $\delta E$ of the excitation causing the
satellite. We therefore keep ratio of $t$, $U$ and $\epsilon
_{a}-\epsilon
_{b}$ fixed, but vary their magnitude. In this way we can vary $\delta
E$
without varying the magnitude of the satellite in the sudden limit. Eq.
(\ref
{eq:m3}) requires that we vary $V_{sc}(0)$ as we vary $\delta E$ (via
$U$),
e.g., by varying the dielectric constant $\varepsilon $. In some of the
calculations below, however, we do not impose Eq. (\ref{eq:m3}), to be
able
to see the effect of varying $\delta E$ alone. We furthermore vary the
range
$R_{0}$ of the potential. From the definition Eq. (\ref{eq:m4}) it
follows
that this would also vary the strength of the potential. For this reason
we
simultaneously vary the dielectric constant $\varepsilon $ so that
$V_{sc}(0)
$ stays unchanged when $R_{0}$ is changed. Alternatively, we can use the
analytical matrix elements (\ref{eq:ma4}, \ref{eq:ma6}). We can then
easily
vary the length scale by changing $\tilde{R}$ or the strength by
changing $%
\tilde{V}$.

To know roughly what are interesting values for our parameters we use
experimental results for some copper dihalides.\cite{Laan} We estimate
the
relative strength of the satellite to the main peak, and the energy
difference between the peaks. This gives two equations while in our
model
these quantities, $r_{00}$ and $\delta E$, depend on three parameters,
$t,U,$
and $\epsilon _{a}-\epsilon _{b}$. To only have two parameters we
consider
{\it the symmetric case} $\epsilon _{a}=\epsilon -U/2$ and $\epsilon
_{b}=\epsilon $ as shown in Fig.\ref{fig:schematic}. 
In the symmetric case we are restricted to the
shake-down
situation since before the transition the $a$-level is above the
$b$-level, $%
\epsilon _{a}+U-\epsilon _{b}=U/2$, while after the transition the
$a$-level
is below the $b$-level, $\epsilon _{b}-\epsilon _{a}=U/2$. In the
symmetric
case we have $0<\theta =\varphi <\pi /4$, and $r_{00}=\cot ^{2}2\varphi
=U^{2}/(16t^{2})$. Once we know where in the ball-park we have $t$ and
$U$,
we can leave the symmetric case, and also consider, e.g., shake-up cases
when there is no level crossing, $\epsilon _{a}>\epsilon _{b}$. In the
lowest final state the electron essentially stays on level $b$, while
the
transfer of the electron to the level $a$ corresponds to a shake up
satellite. In this case we have $-\pi /4<\varphi <0<\theta <\pi /4$ and
$%
\varphi +\theta <0$.

Our calculations usually take the CuCl$_{2}$ parameters as reference
values. For CuCl$_{2}$ we have $\theta =\varphi =       0.3$, which gives 
$r_{00}=2.1$. Further $\tilde{V}=-0.36$ a.u., $\tilde{E}=0.195$ a.u. 
(with $ \tilde{R}=1.6$ a.u.), and $\delta E=0.237$ a.u., i.e.,
$\tilde{V}/\tilde{E}=-1.85$ and $\delta E/\tilde{V}=-0.66$ 
(see also Table II).

\section{Exact treatment}

\subsection{Time-dependent formulation}

To obtain exact results for model Eq.(\ref{eq:6}), we use a
time-dependent
formulation\cite{time-formalism} and solve the Schr\"odinger equation
for
the Hamiltonian
\begin{equation}  \label{eq:6a}
{\cal H}(\tau)={\cal H}_0+T+V+\Delta f(\tau).
\end{equation}
The interaction is switched on at $\tau=0$, using
\begin{equation}  \label{eq:4}
f(\tau)=e^{-i\omega \tau}(e^{-\eta \tau}-1) \hskip1cm \eta>0.
\end{equation}
Here $\eta$ is a small quantity to assure that the external field is
switched on smoothly. The initial ($\tau=0$) state $|\Psi_0\rangle$ is
given
by the ground state of ${\cal H}_0$ with $n_c=1$ in Eq.(\ref{eq:1}).
After a
time $\tau$, the state $|\Psi(\tau)\rangle$ of the system is
\begin{eqnarray}
|\Psi(\tau)\rangle&=&a(\tau)|\psi_a\rangle|\psi_c\rangle
+b(\tau)|\psi_b\rangle|\psi_c\rangle  \nonumber \\
&+&\sum_k c_{ak}(\tau)|\psi_a\rangle|\psi_k\rangle +\sum_k
c_{bk}(\tau)|\psi_b\rangle|\psi_k\rangle.
\end{eqnarray}
The coefficients of $|\Psi(\tau)\rangle$ can be determined by
\begin{equation}
i\frac{\partial}{\partial \tau}|\Psi(\tau)\rangle={\cal H}
(\tau)|\Psi(\tau)\rangle,
\end{equation}
which gives four differential equations for the four coefficients
$a(\tau)$,
$b(\tau)$, $c_{ak}(\tau)$, and $c_{bk}(\tau)$,
\begin{eqnarray}
i\frac{\partial}{\partial \tau}a(\tau)
&=&(\epsilon_a+U+\epsilon_c)a(\tau)+tb(\tau)  \nonumber \\
&+&\sum_k{V_k^d}^{\ast}(\tau)c_{ak}(\tau),
\end{eqnarray}
\begin{eqnarray}
i\frac{\partial}{\partial \tau}b(\tau)
&=&(\epsilon_b+\epsilon_c)b(\tau)+ta(\tau)  \nonumber \\
&+&\sum_k{V_k^d}^{\ast}(\tau)c_{bk}(\tau),
\end{eqnarray}
\begin{eqnarray}
i\frac{\partial}{\partial \tau}c_{ak}(\tau)
&=&(\epsilon_a+\epsilon_k)c_{ak}(\tau)+tc_{bk}(\tau)  \nonumber \\
&+&V_k^d(\tau)a(\tau),
\end{eqnarray}
\begin{eqnarray}
i\frac{\partial}{\partial \tau}c_{bk}(\tau)
&=&(\epsilon_b+\epsilon_k)c_{bk}(\tau) +tc_{ak}(\tau)+V_k^d(\tau)b(\tau)
\nonumber \\
&+&\sum_{k^{\prime}}V_{kk^{\prime}}c_{bk^{\prime}}(\tau),
\end{eqnarray}
where $V_k^d(\tau)=V_0M_kf(\tau)$ with $V_0$ representing the strength
of
the external field. We solve the equations in the limit when $V_0 \to
0$,
and thus the ratio between $c_{ak}$ and $c_{bk}$ is independent of
$V_0$.
The initial conditions are $a(0)=-\sin\theta$, $b(0)=\cos\theta$, and $%
c_{ak}(0)=c_{bk}(0)=0$. Thus the problem is reduced to solving the
coupled
differential equations, which is done using the Runge-Kutta fourth-order
method.

The photoelectron currents $J_1(\omega)$ and $J_2(\omega)$ corresponding
to
main and satellite lines, respectively, are given by
\begin{eqnarray}  \label{eq:t1}
J_1(\omega)&=&\sum_k|\langle\Psi_f^{1k}|\Psi(\tau)\rangle|^2  \nonumber
\\
&=&\sum_k|\cos\varphi c_{ak}(\tau)-\sin\varphi c_{bk}(\tau)|^2,
\end{eqnarray}
\begin{eqnarray}  \label{eq:t2}
J_2(\omega)&=&\sum_k|\langle\Psi_f^{2k}|\Psi(\tau)\rangle|^2  \nonumber
\\
&=&\sum_k|\sin\varphi c_{ak}(\tau)+\cos\varphi c_{bk}(\tau)|^2,
\end{eqnarray}
where $\tau$ is a sufficiently large time. We let the system evolve for
a
time of the order $1/\eta$ to obtain converged results for a given
finite $%
V_0$. In principle, we should perform the calculation for a few small
values
of $\eta$ and then extrapolate to $\eta=0$ followed by an extrapolation
$%
V_0\to 0$. Here, for simplicity we have performed the calculation for
one
single small value of $V_0$. The calculation was performed for
$\eta=0.1$
eV, 0.08 eV, 0.02 eV, and the results were extrapolated to $\eta=0$
assuming
the $\eta$ dependence $a(\omega)\eta+b(\omega) \eta^2+c(\omega)$. The
error
in this approach occurs primarily for small $\tilde{\omega}(\lesssim 5$
eV),
and it is then less than 5\% in $r(\omega)/r_0(\omega)$.

The approach above gives the relative intensity of the main and
satellite
peaks
\begin{equation}
r(\omega)=\frac{J_2(\omega)}{J_1(\omega)}.
\end{equation}
It can be shown that the formulas (\ref{eq:t1}, \ref{eq:t2}) above give
identical results to the more conventional formulation (\ref{eq:pt3})
below,
by performing derivations of the type made in, e.g., Ref.
\onlinecite{auger}.

\begin{figure}
\vspace*{6cm}
\includegraphics{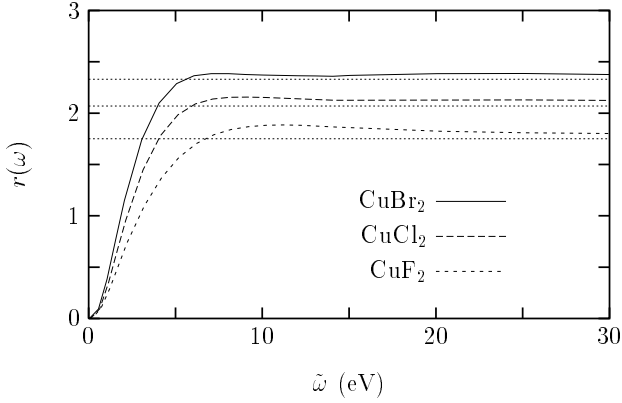}
\caption{The ratio $r(\protect\omega)$ between the satellite and 
the main peak for the divalent copper compounds CuBr$_2$, CuCl$_2$, 
and CuF$_2$.  The parameters are given in Table II. The dotted lines 
are the limit values ($ r(\infty)$) for the respective cases. }
\label{fig:Cu-dihalides}
\end{figure}

As an example of the results obtained in this formalism, we show in Fig.
\ref
{fig:Cu-dihalides} results for the copper dihalides CuBr$_{2}$,
CuCl$_{2}$
and CuF$_{2}$. The corresponding parameters are shown in Table \ref
{used-parameter} and were estimated from experiment.\cite{Laan} The
figure illustrates that there is a small ``overshoot'' for small
$\tilde{\omega}$
but that the sudden limit is reached fairly quickly as $\tilde{\omega}$
is further increased. We remind that in our CuCl$_{2}$ reference case
$\tilde{E}$=0.195 a.u.=5.3 eV.

\subsection{Resolvent formulation}

Alternatively, we can work in the energy space, and obtain the spectrum
by
direct inversion of a resolvent operator. We consider the Hamiltonian
${\cal %
H}$,
\begin{equation}
{\cal H}={\cal H}_{0}+T+V,  \label{eq:pt1}
\end{equation}
where by ${\cal H}_{0}$ we understand ${\cal H}_{0}(n_{c}=0)$. The exact
final photoemission state $|\Psi _{f}^{sk}\rangle $ is\cite{Hedinbook}
\begin{equation}
|\Psi _{f}^{sk}\rangle =\left[ 1+\frac{1}{E-{\cal H}_{0}-T-V-i\eta
}V\right]
|\psi _{s}\rangle |\psi _{k}\rangle ,  \label{eq:pt2}
\end{equation}
where $|\psi _{s}\rangle $ ($s=1,2$) (Eq. (\ref{eq:s4})) are the exact
(target) eigenstates of ${\cal H}_{0}(n_{c}=0)$ and $E=\epsilon
_{k}+E_{s}$
is the energy of the final state. Using Eq. (\ref{eq:pt2}) we calculate
the
matrix element $M(s,{k})\equiv \langle \Psi _{f}^{sk}|\Delta |\Psi
_{0}\rangle $
\begin{eqnarray}
&&M(s,{k})  \label{eq:pt3} \\
&=&\langle \psi _{k}|\langle \psi _{s}|\left[ 1+V\frac{1}{\epsilon
_{k}+E_{s}-{\cal H}_{0}-T-V+i\eta }\right] \Delta |\Psi _{0}\rangle .
\nonumber
\end{eqnarray}
Introducing a basis set
\begin{equation}
|i\rangle =|\psi _{s}\rangle |\psi _{k}\rangle ,  \label{eq:d1}
\end{equation}
the matrix elements of $V$ can then be written as
\begin{equation}
V_{ij}\equiv V_{ks,k^{^{\prime }}s{^{\prime }}}=V_{kk^{^{\prime
}}}v_{s}v_{s^{^{\prime }}}.  \label{eq:d1a}
\end{equation}
Here
\begin{equation}
v_{s}=\cases{-{\rm sin}\varphi,& if $s=1$;\cr {\rm cos}\varphi,& if
$s=2$,\cr}  \label{eq:d1b}
\end{equation}
where we have used Eqs. (\ref{eq:8}, \ref{eq:s4}). The Hamiltonian
matrix
in this basis set is diagonalized, which gives the eigenvalues $\epsilon
_{\nu }$ and the eigenvectors
\begin{equation}
|\nu \rangle =\sum_{i}c_{i}^{\nu }|i\rangle .  \label{eq:d2}
\end{equation}
We then have
\begin{eqnarray}
&&M(s,k)\equiv M(i)=  \label{eq:d3} \\
&&\langle i|\Delta |\Psi _{0}\rangle +\sum_{\nu }\sum_{j,l}{\frac{%
V_{i,j}c_{j}^{\nu }c_{l}^{\nu }\langle l|\Delta |\Psi _{0}\rangle
}{\epsilon
_{k}+E_{s}-\epsilon _{\nu }+i\eta }}.  \nonumber
\end{eqnarray}
The quantities $\langle i|\Delta |\Psi _{0}\rangle $ were given in Eq.
(\ref
{eq:s9}), $\langle i|\Delta |\Psi _{0}\rangle =m_{i}=m_{sk}=M_{k}w_{s}$.
By
organizing the sums in Eq. (\ref{eq:d3}) appropriately, the calculation
of
this expression is very fast and the main time is spent in diagonalizing
the
Hamiltonian matrix. We have found this method to be more efficient than
the
time-dependent method above.

In the expression (\ref{eq:d3}), we can identify the first term as the
intrinsic contribution, since this is the amplitude which is obtained if
there is no interaction between the photoelectron and the target. The
extrinsic effects are then determined by the square of the absolute
value of
the second term. The interference between the intrinsic and extrinsic
contributions is given by the cross product of these terms.

\subsection{Separable potential}

It is interesting to consider a separable potential
\begin{equation}
V_{kk^{^{\prime }}}=\tilde{V}b_{k}b_{k^{^{\prime }}},  \label{eq:sep1}
\end{equation}
since it is then possible to obtain an analytical expression for
$r(\omega )$. The operator in the denominator of Eq. (\ref{eq:pt2}) 
is written as
\begin{equation}
(z-{\cal H}_{0}-T-V)_{ij}=d_{i}(z)\delta _{ij}-\tilde{V}c_{i}c_{j},
\label{eq:sep2}
\end{equation}
where again $|i\rangle =|s\rangle |k\rangle $ is a combined index for
the target state $s$ and the continuum state $k$ and $z$ is a (complex)
number.  Then
\begin{equation}
d_{sk}(z)\equiv d_{i}(z)=z-E_{s}-\epsilon _{k}  \label{eq:sep3}
\end{equation}
and
\begin{equation}
c_{i}\equiv c_{sk}=b_{k}v_{s}.  \label{eq:sep4}
\end{equation}
Using the fact that $V$ is separable, it is then straightforward to
invert the expression in Eq. (\ref{eq:sep2}) and obtain
\begin{eqnarray}
&&[(z-{\cal H}_{0}-T-V)^{-1}]_{ij}  \label{eq:sep5} \\ \nonumber
&=&{\frac{\delta _{ij}}{d_{i}(z)}}+\tilde{V}{\frac{c_{i}c_{j}}{%
d_{i}(z)d_{j}(z)(1-\tilde{V}\sum_{l}c_{l}^{2}/d_{l}(z))}}.
\end{eqnarray}
This leads to
\begin{eqnarray}
&&{\frac{r(\omega )}{r_{0}(\omega )}}=  \label{eq:sep6} \\
&&\left| {\frac{1+{\rm cos}\varphi D_{k_{2}}(E_{0}+\omega )/{\rm cos}%
(\varphi +\theta )}{1+{\rm sin}\varphi D_{k_{1}}(E_{0}+\omega )/{\rm
sin}%
(\varphi +\theta )}}\right| ^{2},  \nonumber
\end{eqnarray}
where $k_{s}$ is defined in Eq.~(\ref{eq:s10}),
\begin{equation}
D_{k}(\epsilon
)=-{\frac{\tilde{V}}{\tilde{E}_{s}}}{\frac{b_{k}E(\epsilon )}{%
M_{k}[1+(\tilde{V}/\tilde{E}_{s})C(\epsilon )]}},  \label{eq:sep7}
\end{equation}
with $\tilde{E}_{s}=1/(2\tilde{R}_{s}^{2})$ and
\begin{equation}
C(\epsilon )=-\tilde{E}_{s}\sum_{l}{\frac{c_{l}^{2}}{d_{l}(\epsilon
+i\eta )}%
}  \label{eq:sep8}
\end{equation}
and
\begin{equation}
E(\epsilon )=-\tilde{E}_{s}\sum_{l}{\frac{c_{l}m_{l}}{d_{l}(\epsilon
+i\eta )%
}}.  \label{eq:sep9}
\end{equation}

To obtain a model for $V_{kk^{^{\prime }}}$ we can, for instance, put
\begin{equation}
b_{k}=\sqrt{\frac{\tilde{R}_{s}}{R}}{\frac{(\tilde{R}_{s}k)^{2}}{1+(\tilde{R}%
_{s}k)^{3}}}.  \label{eq:sep10}
\end{equation}
Compared with the expression in Eq. (\ref{eq:ma6}), there is no term $%
k-k^{^{\prime }}$ in the corresponding expression for $V_{kk^{^{\prime
}}}$
. The neglect of this term means that $V_{kk^{^{\prime }}}$ goes to zero
more slowly as one of the arguments $k$ or $k^{^{\prime }}$ goes to
infinity. To compensate for this we use the power three for
$\tilde{R}_{s}k$
in the denominator of (\ref{eq:sep10}), while in Eq. (\ref{eq:ma6}) the
corresponding power is two. This is a reasonable approximation for small
$k$%
, but it breaks down for large $k$.

\subsection{On the variables in the intensity ratio}

For the satellite to main line intensity ratio we have,
\[
r(\omega)=\frac{k_{1}}{k_{2}} \left|\frac{M(2,k_{2})}{M(1,k_{1})}%
\right|^{2}.
\]
This ratio does not depend on any constant factor in $M_{k}$, since
$M(s,k)$
is proportional to $M_{k}$. If we take the parameters $\tilde{R}_{d}$,
$%
\tilde{R}_{s}$, and $\tilde{R}_{sd}$ equal to a common typical radius $%
\tilde{R}$ (as will be motivated later), and use the analytic
expressions in
Eqs. (\ref{eq:ma4}) and (\ref{eq:ma6}) then $r(\omega)$ or $%
r(\omega)/r_{0}(\omega)$, apart from $\varphi$ and $\theta$, becomes a
function of $\delta E/\tilde{E}$, $\tilde{V}/\tilde{E}$, and
$\tilde{\omega }%
/\tilde{E}$, with $\tilde{E}= (2\tilde{R}^{2})^{-1}$. We can see this
since $%
M_{k}$ is a function of $k\tilde{R}$, and $V_{kk^{\prime }}$ a function
of $k%
\tilde{R}$ and $k^{\prime }\tilde{R}$ apart from their prefactors. The
prefactor of $V_{kk^{\prime}}$ is $\tilde{V}\tilde{R}/R$, while that for
$%
M_{k}$ has no influence. For each $V$ in a perturbation expansion of Eq.
(%
\ref{eq:pt3}) we have an energy denominator and a $k$-summation. The
$k$%
-summation gives an integral and a factor $Rdk.$ Using variables
$\tilde{R}k$%
, the $\tilde{R}/R$ in the prefactor vanishes. Factoring out $\tilde{E}$
in
the energy denominator, we have a factor $\tilde{V}/\tilde{E}$ for each
$%
V_{kk^{\prime}}$, and instead of $\delta E$ and $\tilde{\omega}$ we have
$%
\delta E/\tilde{E}$ and $\tilde{\omega}/\tilde{E}$.

With $\theta $ and $\varphi $ given, $\delta E$ is proportional to $U$.
$U$
in turn is equal to $-V_{sc}(0) $, and thus somehow related to the
strength
of the scattering potential $\tilde{V}.$ If we fix the value of the
sudden
limit $r_{00}=\cot^{2}\left(\varphi+\theta\right)$ by choosing one of
the
angles, we still have an independent parameter left. This parameter can
be
used to decouple the relation between $\delta E$ and $\tilde{V}$
(whatever
it is). Summarizing, we have found that the parameters of our model
system
appear as the angles $\theta $ and $\varphi$, and the excitation energy
$%
\delta E$ (or $U$), while the coupling between the photoelectron and the
model system only appears in one parameter,
$\tilde{V}/\tilde{E}=2\tilde{V}%
\tilde{R}^{2}$, provided we use $\tilde{\omega }/\tilde{E}$ as variable.
We
have further motivated that we can vary the parameters $\delta E$ and $%
\tilde{V}$ independently.

\section{Approximate treatments}

\subsection{Perturbation approach to lowest order in
$V_{\lowercase{kk^{\prime}}}$}
\label{sec:pert}

The same problem can be also studied using the standard perturbation
approach. We consider the expression for the matrix elements $M(s,k)$ in
Eq.  (\ref{eq:pt3}). To lowest order in $V$, we can neglect $V$ in the
denominator of Eq. (\ref{eq:pt3}). Inserting the completeness relation
$ \sum_{i}|i\rangle \langle i|=1$ in terms of eigenstates $|i\rangle
\equiv
|k\rangle |s\rangle $ we obtain
\begin{eqnarray}
&&M(s,k)=\langle s|\langle k|\Delta |\Psi _{0}\rangle   \label{eq:pt0}
\\
&&+\sum_{k^{^{\prime }}s^{^{\prime }}}V_{ks,k^{^{\prime }}s^{^{\prime
}}}[(E-%
{\cal H}_{0}-T+i\eta )^{-1}]_{k^{^{\prime }}s^{^{\prime }},k^{^{\prime
}}s^{^{\prime }}}\langle s^{^{\prime }}|\langle k^{^{\prime }}|\Delta
|\Psi
_{0}\rangle .  \nonumber
\end{eqnarray}
Using Eqs. (\ref{eq:d1a}, \ref{eq:d1b}) we obtain
\begin{eqnarray}
&M&(1,k)=-\sin (\varphi +\theta )M_{k}  \nonumber \\
&-&\sin ^{2}\varphi \sin (\varphi +\theta )\sum_{k^{\prime }}\left[
\frac{%
V_{kk^{\prime }}M_{k^{\prime }}}{E-E_{1}-\epsilon _{k^{\prime }}+i\eta
}%
\right]   \nonumber \\
&-&\frac{\sin 2\varphi \cos (\varphi +\theta )}{2}\sum_{k^{\prime
}}\left[
\frac{V_{kk^{\prime }}M_{k^{\prime }}}{E-E_{2}-\epsilon _{k^{\prime
}}+i\eta
}\right] ,  \label{eq:pt4}
\end{eqnarray}
\begin{eqnarray}
&M&(2,k)=\cos (\varphi +\theta )M_{k}  \nonumber \\
&+&\cos ^{2}\varphi \cos (\varphi +\theta )\sum_{k^{\prime }}\left[
\frac{%
V_{kk^{\prime }}M_{k^{\prime }}}{E-E_{2}-\epsilon _{k^{\prime }}+i\eta
}%
\right]   \nonumber \\
&+&\frac{\sin 2\varphi \sin (\varphi +\theta )}{2}\sum_{k^{\prime
}}\left[
\frac{V_{kk^{\prime }}M_{k^{\prime }}}{E-E_{1}-\epsilon _{k^{\prime
}}+i\eta
}\right] ,  \label{eq:pt5}
\end{eqnarray}
where $V_{kk^{\prime }}=\langle k|V_{sc}|k^{\prime }\rangle $ and $%
M_{k}=\langle k|\Delta |\psi _{c}\rangle $. We can then immediately
calculate the photoemission spectra using Eq.~(\ref{eq:s8}).

\begin{figure}
\vspace*{8.5cm}
\includegraphics{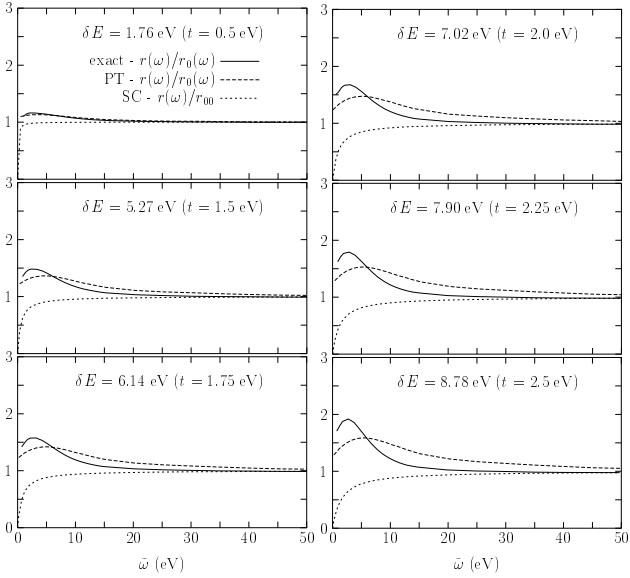}
\caption{$r(\protect\omega)/r_0(\protect\omega)$ from the semi-classical
approximation (SC), the first order perturbation expansion (PT) as well
as
the exact time evolution calculations for different values of the
excitation
energy $\protect\delta E$ and for $U/t=5.76$. The remaining parameters
are
taken from CuCl$_2$ ($R_0^{{\rm Cl}}=4.71$ a.u.). }
\label{fig:ex_energy}
\end{figure}

\begin{figure}
\vspace*{8.5cm}
\includegraphics{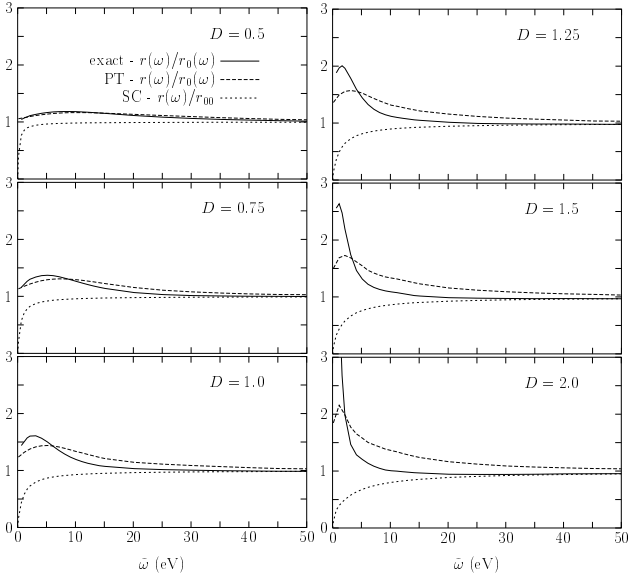}
\caption{The same as in Fig. \ref{fig:ex_energy} but varying the range
$DR_0$
instead of $\protect\delta E$, where $D$ is a scale factor. The
parameters
of CuCl$_2$ are used. }
\label{fig:range}
\end{figure}

In Figs. \ref{fig:ex_energy} and \ref{fig:range} we compare the
perturbation
expansion with the exact time dependent calculation for a realistic
scattering potential in the symmetric case. In the symmetric case we
have $%
\delta E=2t\sqrt{1+r_{00}}= (U/2)\sqrt{1+1/r_{00}}$, and $U=-V_{sc}(0)$.
Since the ratio $r_0(\omega)$ was discussed extensively in Sec. \ref
{sec:sudden}, we here focus on $r(\omega)/r_0(\omega)$, which describes
the
effect of the scattering potential. We vary the excitation energy
$\delta E$
by varying $t$, while keeping $r_{00}$ constant. We also vary the
potential
range by replacing $R_0$ in Eq.~(\ref{eq:m4}) by $DR_0$ and then varying
$D$%
. With $r_{00}$ fixed, $\delta E$ is proportional to $V_{sc}(0)$. Thus a
small $\delta E$ and a small $D$ make the perturbation weak. The
calculations are made for a range of parameter values around those given
for CuCl$_2$ in Table \ref{used-parameter}.

\subsection{Semi-classical approach}

\label{sec:semi}

We can also perform the photoemission calculation by assuming a
classical
trajectory of the emitted photoelectron\cite{Mahan}, producing a
time-dependent potential which drives the dynamics of the the model. It
has been reported that the semi-classical approach can give the 
unexpectedly good results for the systems with coupling to 
plasmons.\cite{Lars98,Inglesfield,Bardy} The essence of the 
semi-classical approach is to replace the scattering potential 
$V_{sc}(r)$ by a time-dependent potential using the charge density 
$\rho ({\bf r},\tau )$ of the emitted electron, {\it i.e.}
\begin{equation}
V_{sc}(r)\rightarrow \int d{\bf r}V_{sc}(r)\rho ({\bf r},\tau
)=V_{sc}(v\tau
),  \label{eq:c1}
\end{equation}
where we have used $\rho ({\bf r},\tau )=\delta ({\bf r}-{\bf v}\tau )$.
We
can then write the Hamiltonian as
\begin{equation}
{\cal H}(\tau )={\cal H}_{0}(n_{c}=0)+V(\tau ),  \label{eq:c2}
\end{equation}
where ${\cal H}_{0}(n_{c}=0)$ can be expressed in terms of the exact
final
states ($\psi _{1}$ and $\psi _{2}$) in the presence of a core hole
\begin{equation}
{\cal H}_{0}(n_{c}=0)=E_{1}\psi _{1}^{\dagger }\psi _{1}+E_{2}\psi
_{2}^{\dagger }\psi _{2}.  \label{eq:c3}
\end{equation}
The time-dependent potential takes the form
\begin{eqnarray}\label{eq:c4}
&&V(\tau )=n_{b}V_{sc}(v\tau )  \nonumber  \\
&=&V_{11}(\tau )\psi _{1}^{\dagger }\psi _{1}+V_{22}(\tau )\psi
_{2}^{\dagger }\psi _{2}+V_{12}(\tau )(\psi _{1}^{\dagger }\psi
_{2}+\psi
_{2}^{\dagger }\psi _{1}),  \nonumber \\
&&
\end{eqnarray}
where (cf.~Eq. (\ref{eq:d1b}))
\begin{eqnarray} \label{eq:c5}
V_{11}(\tau ) &=&\sin ^{2}\varphi V_{sc}(v\tau ),\hskip0.3cmV_{22}(\tau
)=\cos ^{2}\varphi V_{sc}(v\tau ),  \nonumber   \\
V_{12}(\tau ) &=&V_{21}(\tau )=-{\frac{1}{2}}\sin 2\varphi V_{sc}(v\tau
).
\end{eqnarray}
The remaining system (target) is still purely quantum mechanical, and we
write its time-dependent wave function $|\Psi (\tau )\rangle $ as
\begin{equation}
|\Psi (\tau )\rangle =a_{1v}(\tau )|\psi _{1}\rangle e^{-iE_{1}\tau
}+a_{2v}(\tau )|\psi _{2}\rangle e^{-iE_{2}\tau }.  \label{eq:c6}
\end{equation}
The classical electron velocity $v$ is determined by energy
conservation,
that is, $\frac{1}{2}v^{2}=\tilde{\omega}$. We have here chosen the
velocity
corresponding to the satellite. We could also have performed two
calculations, with the velocities corresponding to the leading peak and
to
the satellite, respectively. In contrast to the approach used here,
this would, however, lead to the problem that the spectral weight 
would not be normalized. Applying the time-dependent Schr\"{o}dinger 
equation to $|\Psi (\tau )\rangle $, we obtain $a_{1v}(\tau )$ and 
$a_{2v}(\tau )$,
\begin{equation}
i\frac{\partial }{\partial \tau }a_{1v}(\tau )=V_{11}(\tau )a_{1v}(\tau
)+V_{12}(\tau )a_{2v}(\tau )e^{-i\delta E\tau },  \label{eq:c7}
\end{equation}
\begin{equation}
i\frac{\partial }{\partial \tau }a_{2v}(\tau )=V_{22}(\tau )a_{2v}(\tau
)+V_{21}(\tau )a_{1v}(\tau )e^{i\delta E\tau },  \label{eq:c8}
\end{equation}
where $\delta E$ (Eq.~(\ref{eq:s7})) is the optical excitation energy.
Eqs.~(%
\ref{eq:c7}) and (\ref{eq:c8}) are subject to the initial conditions 
(cf Eq. \ref{eq:s9})
\begin{equation}
a_{sv}(0)=w_{s}.
\end{equation}
The final photoemission currents $J_{i}(\omega )$ is
\begin{equation}
J_{i}(\omega )\propto |a_{iv}(\tau _{0})|^{2},\mbox{ }\mbox{ }i=1,2.
\end{equation}
It is sufficient to perform the calculation up to $\tau =\tau _{0}\equiv
R_{0}/v$, since the potential vanishes for larger values of $\tau $. The
relative intensity between main and satellite contributions is given by
$%
r(\omega )=J_{2}(\omega )/J_{1}(\omega )$ as before.

In Figs. \ref{fig:ex_energy} and \ref{fig:range} we compare the
semi-classical and exact results for a realistic potential in the
symmetric
case. The semi-classical theory is inaccurate over most of the energy
range
considered here. For large energies however, the semi-classical theory
comes
much closer to the exact result than does the PT. It is also clear that
an
increasing $\delta E$ ($\simeq V_{sc}(0)$) does not noticeably affect
the
energy for the adiabatic-sudden transition, where it strongly effects
the
maximum deviation. An increasing $D$ on the other hand not only strongly
increases the maximum deviation, but also makes the adiabatic-sudden
transition energy smaller. The dependence on the parameters will be
investigated more extensively in the next section.

\section{Adiabatic-sudden transition}

\label{sec:adiabatic}

We are now in a position to address the adiabatic-sudden transition and
its
dependence on the parameters. The calculations are performed with the
analytical matrix elements in Eqs. (\ref{eq:ma4}, \ref{eq:ma6}). First
we
study the different characteristic lengths, $\tilde{R}_{d}$ for the
dipole
matrix elements, and $\tilde{R}_{s}$ and $\tilde{R}_{sd}$ for the
scattering
potential matrix elements. We find that it makes sense to use only one
effective length $\tilde{R}$, and the corresponding energy
$\tilde{E}=1/(2\tilde{R}^{2})$.
As we remarked in Sec. VD, $r/r_{0}$ as a function of $%
\tilde{\omega}/\tilde{E}$ depends on the parameters $\delta E/\tilde{E}$
and
$\tilde{V}/\tilde{E}$, and also on the ''system'' parameters $\theta $
and $%
\varphi $. We vary $\delta E$ independently of $\tilde{V}$, although for
a
given model there is a direct relation between these two quantities.
Part of
this relation can be offset by using different $\theta $ and $\varphi $
(with $r_{00}$ constant) but we do not explore this possibility. The
exact
solution with a separable potential is used to discuss the validity and
breakdown of perturbation theory. We find that $\tilde{V}/\tilde{E}$
has a large effect on the deviation from the sudden limit, but little 
effect on the value of $\tilde{\omega}/\tilde{E}$ where the deviation 
becomes small,
while $\delta E/\tilde{E}$ has a comparatively small effect on both
magnitude and range of the deviation. For simplicity we use the 
CuCl$_{2}$ parameters $\theta=\varphi=       0.3$, which gives $r_{00}=2.1$.
For CuCl$_2$ we further have $\tilde{V}$=-0.36 a.u., 
$\tilde{E}$=0.195 a.u. ($\tilde{R}$=1.6 a.u.), and $\delta E=0.237$ 
a.u., i.e., $\tilde V/\tilde E=-1.85$ and $\delta E/\tilde V=-0.66$.
In our calculations, we vary $\tilde V/\tilde E$ and $\delta E/\tilde V$ 
by typically a factor of two around these reference values.

\subsection{Exact numerical treatment with analytic matrix elements}

\begin{figure}
\centerline{\rotatebox{-90}{\epsfysize=3.5in
\epsffile{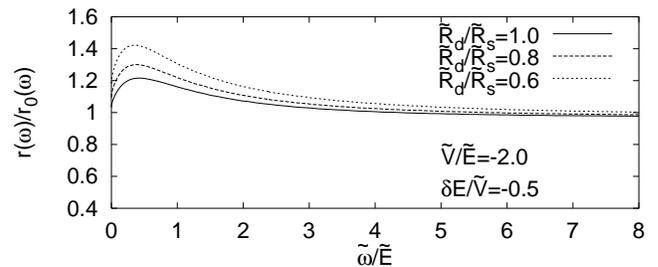}}}
\caption{The ratio $r(\protect\omega)/r_0(\protect\omega)$ as a function
of $ \tilde R_d/\tilde R_s$ for a fixed $\tilde R_{sd}/\tilde R_s=1$ 
and for $\protect\varphi=\protect\theta=0.3$. The figure illustrates 
that there are no qualitative changes as the length scales for the 
dipole and scattering matrix elements become different.}
\label{fig:rd}
\end{figure}

We first illustrate the dependence on the ratio between the length
scales $ \tilde{R}_{d}$, $\tilde{R}_{s},$ and $\tilde{R}_{sd}$. In Fig.
\ref{fig:rd}
we show the results for different ratios $\tilde{R}_{d}/\tilde{R}_{s}$
keeping $\tilde{R}_{sd}/\tilde{R}_{d}=1$. These results are obtained for
$%
\tilde{V}/\tilde{E}_{s}=-2.0$ and $\delta E/\tilde{V}=-0.5$, where
$\tilde{E}%
_{s}=1/(2\tilde{R}_{s}^{2})$ is the energy scale set by the scattering
potential length scale. Fig. \ref{fig:rd} shows that as $\tilde{R}_{d}/%
\tilde{R}_{s}$ is reduced the magnitude of the "overshoot" is increased.
There are, however, no qualitative changes.

\begin{figure}
\centerline{\rotatebox{-90}{\epsfysize=3.5in
\epsffile{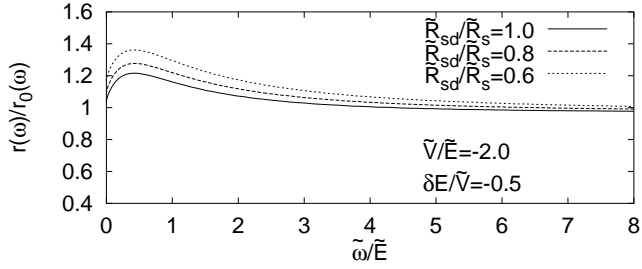}}}
\caption{The ratio $r(\protect\omega)/r_0(\protect\omega)$ as 
a function of $\tilde R_{sd}/\tilde R_s$ for a fixed $\tilde 
R_{d}/\tilde R_s=1$ and for $\protect\varphi=\protect\theta=0.3$. 
The figure illustrates that there is no qualitative changes as 
the ratio of the two length scales in the scattering matrix 
elements is varied.}
\label{fig:rsd}
\end{figure}

Fig. \ref{fig:rsd} shows results for different values of
$\tilde{R}_{sd}/ \tilde{R}_{s}$ for a fixed $\tilde{R}_{d}/
\tilde{R}_{s}=1$. From Eq.  (\ref {eq:ma6}) it can been seen 
that this corresponds to varying the range of values $k-k^{^{\prime }}$ 
where $V_{kk^{^{\prime }}}$ is large, without changing the range 
over which $V_{kk}$ varies. The figure illustrates that the overshoot 
becomes larger as $\tilde{R}_{sd}/\tilde{R}_{s}$ is reduced.
This is natural, since decreasing $\tilde{R}_{sd}$ effectively makes the
scattering potential stronger by expanding the range of values $
k-k^{^{\prime }}$ with large scattering matrix elements. The qualitative
behaviour, however, is not changed. In view of Figs. \ref{fig:rd} and
\ref {fig:rsd}, we study below the case when $\tilde{R}_{sd}=
\tilde{R}_{s}=\tilde{ R}_{d}$, as mentioned in Sec. VD.

Fig. \ref{fig:strength} shows such results for different values of the
strength of the scattering potential $\tilde V/\tilde E$ and for
different values of the excitation energy $\delta E/\tilde V$. 
In each panel $\delta E /\tilde V$ is kept fixed, but the ratio 
is varied by a factor of four from Fig. \ref{fig:strength}a to
 Fig. \ref{fig:strength}c. Typically $r(\omega)/r_0(\omega)$ has   
an overshoot for small values of $\tilde{\omega}$. For somewhat 
larger $\omega$ the ratio approaches unity and possibly becomes 
smaller than unity. The overshoot can be fairly large and happens 
on a small energy scale ($\sim \tilde E$).  In a
few cases of a large overshoot, $r(\omega)/r_0(\omega)$ does not become
approximately unity until $\tilde{\omega}$ is several times $\tilde E$,
although the relevant energy scale is still $\tilde E$. In the case of
an
undershoot, $r(\omega)/r_0(\omega)$ approaches unity from below very
slowly
(energy scale much larger than $\tilde E$). The undershoot is, however,
relatively small, and if we do not require a high accuracy, we consider
the
sudden approximation is valid when the overshoot becomes small. This
means
that as the range of the scattering potential is made larger, the sudden
limit is reached at a smaller energy. This is the opposite to what one
would
expect from the semi-classical theory. The figure illustrates that
$\delta E$
is not the relevant energy scale. Since in each panel we keep $\delta
E/%
\tilde V$ fixed, there is a variation of $\delta E/\tilde{E}$ by a
factor of
four. Furthermore there is a variation of $\delta E/\tilde{V}$ by a
factor
of four in going from the top to the bottom panel in Fig.
\ref{fig:strength}. There is no corresponding change in the energy 
for the adiabatic to sudden transition.

\subsection{Separable potential}

\begin{figure}
\centerline{\rotatebox{-90}{\epsfxsize=2.5in
\epsffile{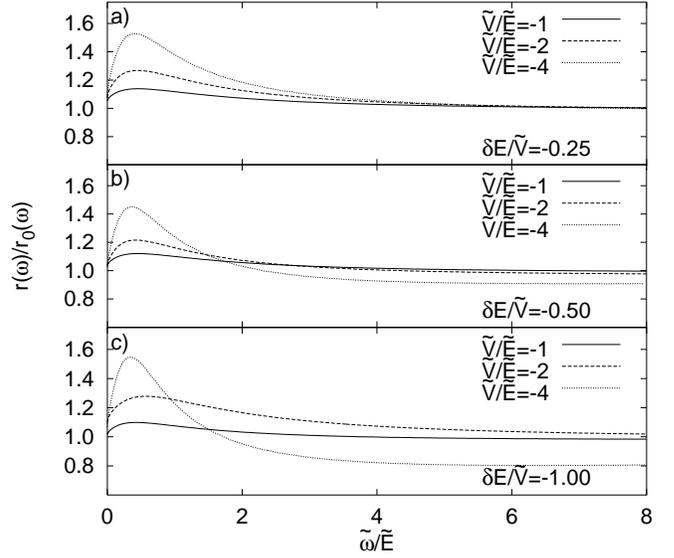}}}
\vskip0.8cm
\caption{The ratio $r(\protect\omega )/r_{0}(\protect\omega )$ as a
function
of $\tilde{\protect\omega}/\tilde{E}$ for different values of
$\tilde{V}/%
\tilde{E}$ and $\protect\delta E/\tilde{V}$ and for $\protect\varphi =%
\protect\theta =0.3$. The figure illustrates that $\tilde{E}$ is an
appropriate energy scale for the adiabatic to sudden transition.}
\label{fig:strength}
\end{figure}

\begin{figure}
\vspace*{9.5cm}
\includegraphics{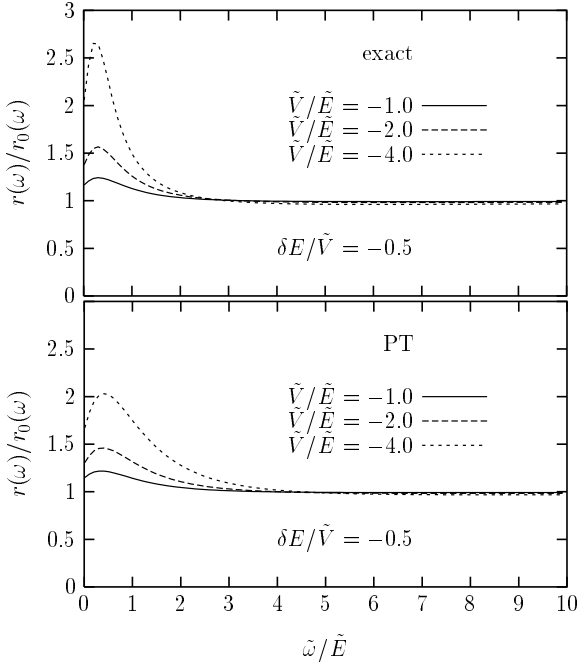}
\caption{The ratio $r(\protect\omega)/r_0(\protect\omega)$ for a
separable scattering potential (\ref{eq:sep1}). According to the 
exact result in upper
panel, the overshoot is substantially overestimated by the separable
potential compared to Fig. \ref{fig:strength}b, but there is still a
qualitative agreement with the more realistic model (\ref{eq:ma6}). The
lower panel shows how perturbation theory works and it illustrates the
effects of multiple scattering. }
\label{fig:separable}
\end{figure}

It is interesting to study a separable potential, since it is then
possible
to obtain an analytical solution. This makes it easier to interpret the
results. It also allows the study the effects of multiple scattering,
i.e.  the deviations from first order perturbation theory. Fig.
\ref{fig:separable}
shows results of the exact and first order theory using the same values
of $ \delta E/\tilde{V}$ and $\tilde{V}/\tilde{E}$ as in Fig.
\ref{fig:strength}b. The separable potential 
overestimates the magnitude of the overshoot in $r(\omega )/
r_{0}(\omega)$ quite substantially. Otherwise the results are rather 
similar. For a qualitative
discussion, we can therefore use the separable potential.

\begin{figure}
\vspace*{9.5cm}
\includegraphics{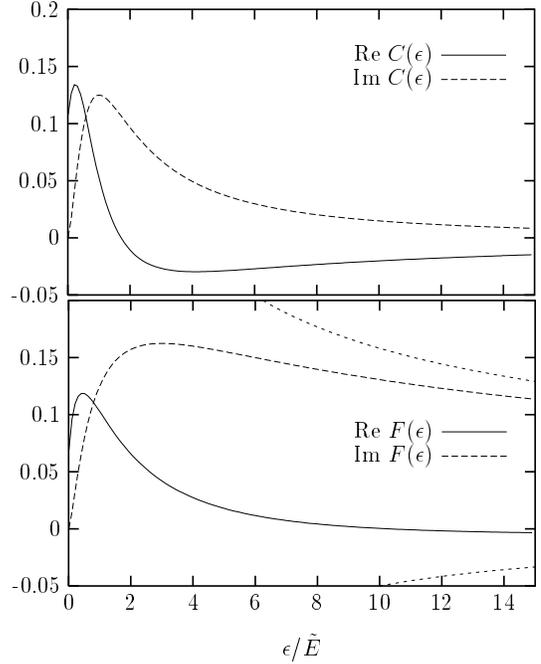}
\caption{The function $C(\epsilon)$ relevant for a separable
potential is given in the upper panel. The low panel gives the 
behaviors of $F(\epsilon)$ defined in Eq.(\ref{eq:Fk1}). 
The dotted lines give the asymptotic behaviors of $F(\epsilon)$
in Eq.(\ref{eq:apt8}).  }
\label{fig:universal}
\end{figure}

For simplicity, we consider $\delta E=0$. We further put
$M_{k}=(\tilde{R}k)^{2}/\left[ 1+(\tilde{R}k)^{3}\right] =b_{k}$. 
This is a poor approximation for large $k$, but then anyhow also 
$V_{kk^{\prime }}$ is poorly represented by the separable potential. 
Our approximations lead to simple results for the functions $C$, 
$D$ and $E$ entering in Eqs.  (\ref{eq:sep6}-\ref{eq:sep10}).
\begin{equation}
C(\epsilon )=-\tilde{E}\sum_{k^{^{\prime }}}{\frac{b_{k^{^{\prime
}}}^{2}}{%
\epsilon -\epsilon _{k^{^{\prime }}}+i\eta }}  \label{eq:as1}
\end{equation}
and $E(\epsilon )={\rm cos}\theta \tilde{M}C(\epsilon )$. Then
\begin{equation}
D(\epsilon )=-{\rm cos}\theta
{\frac{\tilde{V}}{\tilde{E}}}{\frac{C(\epsilon
)}{1+(\tilde{V}/\tilde{E})C(\epsilon )}}.  \label{eq:as2}
\end{equation}
Since $D_{k}$ is independent of $k$ in this approximation, we have
dropped
the index $k$. The function $C(\epsilon )$ is shown in Fig. \ref
{fig:universal}. For $\varphi =\theta $ we can then write
\begin{equation}
{\frac{r(\omega )}{r_{0}(\omega )}}=\left| {\frac{1-\frac{{\rm cos}%
^{2}\varphi }{{\rm cos}2\varphi
}\frac{\tilde{V}}{\tilde{E}}C(\tilde{\omega}%
)/[1+\frac{\tilde{V}}{\tilde{E}}C(\tilde{\omega})]}{1-\frac{1}{2}\frac{%
\tilde{V}}{\tilde{E}}C(\tilde{\omega})/[1+\frac{\tilde{V}}{\tilde{E}}C(%
\tilde{\omega})]}}\right| ^{2}.  \label{eq:as3}
\end{equation}
For a level ordering as indicated in Fig. \ref{fig:schematic},
$0<\varphi
<\pi /4$ and cos$^{2}\varphi /{\rm cos}(2\varphi )\geq 1$. In our
standard
case with $\varphi =0.3$, $\cos ^{2}\varphi /\cos (2\varphi )=1.11$.
Thus
the term in the numerator of Eq. (\ref{eq:as3}) dominates. The factor
$1+(%
\tilde{V}/\tilde{E})C(\tilde{\omega})$ gives the multiple scattering,
which
is not included in the first order perturbation theory, i.e. the first
order
result is
\begin{equation}
\left[ \frac{r(\omega )}{r_{0}(\omega )}\right] _{{\rm PT}}=\left|
\frac{1-%
\frac{\cos ^{2}\varphi }{\cos 2\varphi
}\frac{\tilde{V}}{\tilde{E}}C(\tilde{%
\omega})}{1-\frac{1}{2}\frac{\tilde{V}}{\tilde{E}}C(\tilde{\omega})}\right|
^{2}.  \label{eq:as3a}
\end{equation}
We now compare the behaviors of Eqs. (\ref{eq:as3}) and (\ref{eq:as3a}),
to see the effects of using perturbation theory. For small values of
$\tilde{\omega}$, Re$C(\tilde{\omega})$ is positive and then changes 
sign at about $\tilde{ \omega}/\tilde{E}\sim 2$. Im $C(\tilde{\omega})$ 
is always positive. Due to
our crude approximations for $b_{k}$ and $M_{k}$, $C(\tilde{\omega})$
rapidly becomes unreliable beyond $\tilde{\omega}/\tilde{E}=2$. Both for
the exact and perturbative expressions $r/r_{0}$ goes from over- to
undershoot approximately when Re$C(\tilde{\omega})=0$. This is somewhat 
earlier than in Fig. 11, where however $\delta E=0.5$. 
Comparing $r(\omega)/r_{0}(\omega )$
for the exact (Eq. (\ref{eq:as3})) and the first order result (Eq. (\ref
{eq:as3a})), we find that the exact solution is larger when ${\rm Re}C(%
\tilde{\omega})\gtrsim {\rm Im}C(\tilde{\omega})$, cf Fig. \ref
{fig:separable}. This is consistent with second order perturbation
theory, which is found to enhance $r(\omega )/r_{0}(\omega )$ for small
$\tilde{ \omega}$ and reduce it for large $\tilde{\omega}$. For
$\tilde{\omega}=0$, $C(\tilde{\omega})$ is purely real and slightly 
larger than 0.1, thus multiple scattering gives a divergence in both 
$J_{1}(\tilde{\omega})$ and $J_{2}(\tilde{\omega})$ when 
$\tilde{V}/\tilde{E}\sim -10$. This is due to the attractive potential
$V$ forming a bound state from the continuum states.

The important conclusion from analysing the separable potential is 
that if $\tilde V$ is not too large, first order perturbation 
theory gives roughly the correct range over which there are essential 
deviations from the sudden limit, while multiple scattering
increases the magnitude of these deviations for small $\tilde{\omega}$
and slightly decreases them for larger $\tilde{\omega}$.

We next discuss the physical interpretation of the expression 
(\ref{eq:as3}) (or the perturbational expressions in (\ref{eq:pt0},
\ref{eq:pt4}, \ref{eq:pt5})). Here unity (the first term in
(\ref{eq:pt0}, \ref{eq:pt4}, \ref{eq:pt5})) corresponds to 
a direct transition into the final continuum state 
corresponding to energy conservation. The second term  
(the last two terms in (\ref{eq:pt0}, \ref{eq:pt4}, \ref{eq:pt5}))
corresponds to a virtual transition into some other 
continuum state followed by one or several scattering events 
with the electron ending up in the continuum state corresponding
to energy conservation.  Let us consider 
the virtual emission into a continuum state with a larger 
energy than the final state and let this be followed by one 
scattering event into the final state. For a negative $\tilde V$
the interference with the direct event is then constructive.
For small photon energies such events dominate for two reasons. 
Firstly, there are many more states available above the energy 
corresponding to energy conservation than below, and secondly 
the dipole matrix elements suppress the transitions to the 
energies below. As a result, both the main peak and the satellite 
are enhanced by the scattering effects. For the values of $\varphi$ 
and $\theta$ considered here ($<\pi/4$), the relative effect is 
stronger for the satellite. As a result $r(\omega)/r_0(\omega)$
is enhanced. For larger photon energies Re$C(\tilde{\omega})$
becomes negative. The density of states of partial waves 
with given $l$ and $m$ quantum numbers decreases with 
energy ($\sim 1/\sqrt{\epsilon})$). This favours virtual 
emissions to states below the final continuum state. Depending
on the model for $b_k$ and $M_k$, these matrix elements
may have the same effect. As a result, Re $C(\tilde{\omega})$ becomes
slightly negative for large energies and the ratio 
$r(\omega)/r_0(\omega)$ is slightly smaller than one. 

The relevant energy scale for $C(\omega)$ is $\tilde E$. This is a 
combination of the two effects discussed above. The turn on of
the dipole matrix elements on an energy scale of the order $\tilde E$
favors an increasing value of $C$ over this energy scale, while the density
of states effects becomes more important for larger energies. As a result,
both Re $C(\omega)$ and Im $C(\omega)$ have a maximum at an energy
of the order $\tilde E$. 

\subsection{Perturbational treatment with analytic matrix elements}

In this section we study the perturbation theory expression in more
detail and without relying on a separable potential. Instead we 
consider the more realistic matrix elements in Eqs. (\ref{eq:ma4}) 
and (\ref{eq:ma6}), assuming that $\tilde{R}_{d}=\tilde{R}_{s}=
\tilde{R}_{sd}=\tilde{R}$. We define a function $F_{k}$ by
\begin{equation}
\sum_{k^{^{\prime }}}{\frac{V_{kk^{^{\prime }}}M_{k^{^{\prime
}}}}{\epsilon
-\epsilon _{k^{^{\prime }}}+i\eta }}\equiv
-{\frac{\tilde{V}}{\tilde{E}}}%
M_{k}F_{k}(\epsilon /\tilde{E}),  \label{eq:Fk1}
\end{equation}
which is possible due to the simple form of $V_{kk^{^{\prime }}}$ and $%
M_{k^{^{\prime }}}$. Explicitly we have
\begin{equation}
F_{k}\left( \epsilon \right) =\frac{1}{\pi }\int_{0}^{\infty }\frac{%
x^{4}dx}{\left[ 1+x^{2}\right] ^{2}\left[ 1+\left(
\tilde{R}k-x\right)
^{2}\right] \left[ x^{2}-\epsilon -i\eta \right] }.  \label{eq:Fk2}
\end{equation}
 From Eqs. \ref{eq:pt4} and \ref{eq:pt5} we have
\begin{eqnarray}\label{eq:as4}
& &\frac{r(\omega )}{r_{0}(\omega )}=
\\ \nonumber
& &\left| \frac{%
1-\cos ^{2}\varphi
{\frac{\tilde{V}}{\tilde{E}}}F_{k_{2}}\left(\frac{\tilde{\omega }}%
{\tilde{E}}\right)-\frac{\sin 2\varphi \sin (\varphi +\theta )}{2\cos (\varphi
+\theta )}{\frac{\tilde{V}}{\tilde{E}}}F_{k_{2}}\left(\frac{
\tilde{\omega }
+\delta E}{\tilde{E}}\right)}{1-\sin ^{2}\varphi
{\frac{\tilde{V}}{\tilde{E}}%
}F_{k_{1}}\left(\frac{\tilde{\omega }+\delta E}
{\tilde{E}}\right)-\frac{\sin
2\varphi \cos (\varphi +\theta )}{2\sin (\varphi +\theta
)}{\frac{\tilde{V}}{ \tilde{E}}}F_{k_{1}}\left(\frac{\tilde{\omega }}
{\tilde{E}}\right)}\right| ^{2}.
\end{eqnarray}
Since $\tilde{R}k_{2}=\sqrt{\tilde{\omega }/\tilde{E}}$ and $%
\tilde{R}k_{1}=\sqrt{\left( \tilde{\omega }+\delta E\right)
/\tilde{E%
}}$ we see, as stated earlier, that $r/r_{0}$ depends only on
$\tilde{V}/%
\tilde{E}$, $\delta E/\tilde{E}$ and $\tilde{\omega
}/\tilde{E}.$

First we consider the limit of small values of $k$ and $\delta E$. For
$ \tilde{\omega }=\delta E=0$ we have $F_{0}\left( 0\right) =1/16$,
and
\begin{equation}
\frac{r(0)}{r_{0}(0)}=\left[ {\frac{1+\frac{|\tilde{V}|}{\tilde{E}%
}\cos \varphi \cos \theta /(16\cos (\varphi +\theta ))}{1+\frac{
|%
\tilde{V}|}{\tilde{E}} \sin \varphi \cos \theta /(16\sin
(\varphi
+\theta ))}}\right] ^{2}.  \label{eq:apt7}
\end{equation}
If the more localized level $a$ is above $b$ (see Fig.
\ref{fig:schematic})
in the initial state and below $b$ in the final state (``shake-down''),
we have $0<\varphi <\pi /4$ and $0<\theta <\pi /4$, and the factor in 
the brackets is larger than unity. Thus interaction effects enhance 
the ratio $r(\omega )/r_{0}(\omega )$. This corresponds to a constructive
interference between intrinsic and extrinsic effects. This is in 
contrast to the destructive interference found for the plasmon
case.\cite{Lars98,plasmon} The present treatment, however, refers to 
the ``shake-down'' scenario, and
it is more appropriate to compare the plasmon case with the ''shake up''
case ($-\pi /4<\varphi <0<\theta <\pi /4$ and $\varphi +\theta <0$).
Then
the expression (\ref{eq:apt7}) for $r(\omega )/r_{0}(\omega )$ indeed
becomes smaller than one, and the relative weight of the satellite is
reduced for small energies. We notice, however, that both the satellite
and the main peak are enhanced by the interference, but that the main peak
is enhanced more in the ''shake-up'' situation.

We next consider the case when $k$ is large. 
$F_{k}\left(
\epsilon \right) $ for large $k$ and $\epsilon $ is (with
$\tilde{R }k\approx \sqrt{\epsilon }$)

\begin{equation}
F_{k}\left( \epsilon \right) =\frac{1}{2\sqrt{\epsilon }}\left[
i- \frac{1}{2\sqrt{\epsilon }}+\tilde{R}k-\sqrt{\epsilon}\right].
\label{eq:apt8}
\end{equation}
For the case when $\theta =\varphi $ we obtain
\begin{equation}
\frac{r(\omega )}{r_{0}(\omega)}-1=-{\frac{|\tilde{V
}|}{2\tilde{\omega }}\frac{1+\delta E/\tilde{E}}{2\cos (2\varphi
)}+ \frac{|\tilde{V}|^{2}}{4\tilde{\omega }\tilde{E}}}\left(
{\frac{\cos^{4}\varphi }{\cos^{2}(2\varphi )}}-{\frac{1}{4}}\right).  
\label{eq:apt8a}
\end{equation}
Thus the approach to the sudden limit goes as $1/\tilde{\omega }$
with a coefficient which depends on the parameters. With our standard
CuCl$_{2}$ parameters, we have
\[
\frac{r(\omega )}{r_{0}(\omega )}-1=-0.30{\frac{|%
\tilde{V}|}{\tilde{\omega }}}\left( 1+\frac{\delta
E}{\tilde{E}}%
\right) +0.24{\frac{|\tilde{V}|^{2}}{\tilde{\omega }\tilde{E}}}
\]
We note that for large $|\tilde{V}|/\tilde{E}$, and when
$|\tilde{V}|$ is large enough compared to $\delta E,$ the last 
(positive) term dominates.  The approach to the sudden limit is 
then set by $|\tilde{V}|^{2}/\tilde{E }=2\left( \tilde{R}
\tilde{V}\right) ^{2}$. 

To evaluate Eq.(\ref{eq:as4}) when $\delta E=0$ we only need the function
$F(\epsilon)$,
$$
F\left( \epsilon \right) =\frac{1}{\pi }\int_{0}^{\infty }\frac{%
x^{4}dx}{\left[ 1+x^{2}\right] ^{2}\left[ 1+\left(
\sqrt{\epsilon}-x\right)
^{2}\right] \left[ x^{2}-\epsilon -i\eta \right] }.
$$
We show $F\left( \epsilon \right) $ in the lower panel of 
Fig.\ref{fig:universal}. The
results Eq. (\ref{eq:apt8}) for large values of $\epsilon $ are shown by
the dotted lines in the figure. Clearly the approach of Re $%
F$ to its asymptote is very slow. If we take $\delta E=0$
we have the same form as in first order perturbation theory
with a separable potential Eq. (\ref{eq:as3a}),
\begin{equation}
\frac{r(\omega )}{r_{0}(\omega )}=\left| \frac{1-\frac{\cos ^{2}\varphi
}{%
\cos 2\varphi
}\frac{\tilde{V}}{\tilde{E}}F(\tilde{\omega})}{1-\frac{1}{2}%
\frac{\tilde{V}}{\tilde{E}}F(\tilde{\omega})}\right| ^{2},
\label{eq:apt100}
\end{equation}
where we have put $\varphi=\theta$.
As shown in Fig. \ref{fig:universal}, $F(\epsilon )$ has a qualitatively
similar behavior as $C(\epsilon )$ for $\epsilon/\tilde E\lesssim 2$. 
As in the case of the function $C$, the relevant energy scale is
$\tilde E$. 

\begin{figure}
\vspace*{1.cm}
\centerline{\rotatebox{-90}{\epsfxsize=2.5in \epsffile{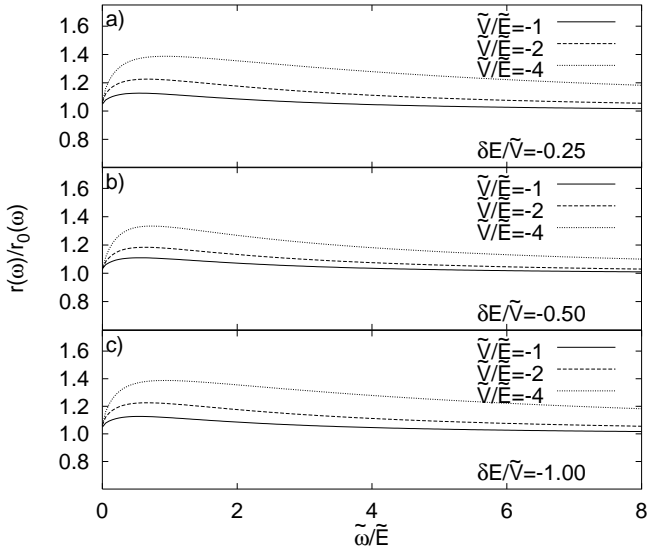}}}
\vskip0.8cm
\caption[]{\label{fig:pert}
The ratio $r(\omega)/r_0(\omega)$ as a function of
$\tilde{\omega}/\tilde E$ for different values of
$\tilde V/\tilde E$ and $\delta E/\tilde V$.}
\end{figure}

In Fig. \ref{fig:pert} we show $r(\omega )/r_{0}(\omega )$ as a function
of $%
\tilde{\omega}/\tilde{E}$ for a few values of $\delta E/\tilde{V}$ and
$%
\tilde{V}/\tilde{E}$. We see that $r(\omega )/r_{0}(\omega )$ starts at
a
positive value (cf Eq.~(\ref{eq:apt7})), and reaches a broad maximum at
about $\tilde{\omega}/\tilde{E}\sim 0.5-1.5$. Compared to the separable
potential solution in Fig. \ref{fig:separable}, the overshoot behavior
is
robust up to fairly large energies, which is due to
Im$F(\tilde{\omega})$
decaying more slowly than Im$C(\tilde{\omega})$. For much larger values
of $%
\omega ,$ $\left( r/r_{0}-1\right) $ decays as $1/\tilde{\omega}$, as
shown
in Eq.~(\ref{eq:apt8a}). Here, based on the discussion in Sec. VIIB, we
can
expect that as multiple scattering becomes important, the region where
there
is an overshoot is substantially reduced and the region with an
undershoot
becomes larger. At the same time, the overshoot intensity will be
enhanced.
These behaviors are actually confirmed by comparing with the exact
calculations given in Fig. \ref{fig:strength}.

Fig. \ref{fig:pert} illustrates that for intermediate values of
$\tilde{%
\omega}$, when Re$F$ dominates, $\left( r/r_{0}-1\right) $ goes as
roughly $%
|\tilde{V}|/\tilde{\omega}$ if $\delta E/\tilde{E}$ is not too large.
For
larger (but not too large) values of $\tilde{\omega}$, Re$F$ becomes
small
and Im$F$ dominates.
Since Re$F$ is positive for small energies, this leads to a constructive
interference between intrinsic and extrinsic effects. For energies of
the order $\tilde{E}$, 
Re$F$ changes sign, and the interference becomes weakly
destructive. For somewhat larger energies the extrinsic effects are
mainly
determined by the imaginary part of $F$. From Eq. (\ref{eq:apt8a}) it
follows
that in perturbation theory the extrinsic effects become small on the
energy
scale $\tilde{V}^{2}/\tilde{E}$.

\subsection{Semi-classical approximation}

In this section we analyze the adiabatic-sudden transition within the
semi-classical framework. From the coupled differential equations
Eqs.~(\ref
{eq:c7}) and (\ref{eq:c8}) we can obtain differential equations for $%
\partial |a_{iv}(\tau )|^{2}/\partial \tau $, $i=1,2$. Integration of
these
equations, leads to
\begin{eqnarray}\label{eq:a1}
&&|a_{2v}(\tau _{0})|^{2}-|a_{2v}(0)|^{2}  \nonumber \\
&=&2{\rm Im}\int_{0}^{\tau _{0}}V_{12}(\tau )a_{1v}(\tau )a_{2v}^{\ast
}(\tau )e^{i\delta E\tau }d\tau ,
\end{eqnarray}
where $\tau _{0}=R_{0}/v$ is the time at which the emitted electron with
the
velocity $v$ leaves the range $R_{0}$ of the scattering potential. $%
|a_{iv}(\tau _{0})|^{2}-|a_{iv}(0)|^{2}$ is a measure of the deviation
from
the sudden limit. For small values of $\tau $ both the coefficients $%
a_{iv}(\tau )$ and the exponent $e^{i\delta E\tau }$ are approximately
real
and there is a small contribution to the imaginary part of the integral
in
Eq.~(\ref{eq:a1}). As $\tau $ grows there is, however, a contribution
from
both these sources.

To obtain a qualitative understanding of the semi-classical
approximation,
we solve the Schr\"{o}dinger equations Eqs.~(\ref{eq:c7}) and
(\ref{eq:c8})
to lowest order in $1/v$. This leads to
\begin{eqnarray}
&&a_{1v}(\tau )=a_{1v}(0)  \label{eq:a1a} \\
&&-i\int_{0}^{\tau }d\tau ^{^{\prime }}[V_{11}(\tau ^{^{\prime
}})a_{1v}(0)+V_{12}(\tau ^{^{\prime }})a_{2v}(0)]  \nonumber
\end{eqnarray}
and a similar result for $a_{2v}(\tau )$. This gives
\begin{eqnarray}
&&|a_{2v}(\tau _{0})|^{2}-|a_{2v}(0)|^{2}={\frac{1}{2}}\sin (2\varphi
)\left\{ {\frac{1}{2}}\sin (2\theta )\right.   \label{eq:a1b} \\
&&\left. \times \left[ \int_{0}^{\tau _{0}}d\tau V_{sc}(\tau )\right]
^{2}+\sin (2\varphi +2\theta )\delta E\int_{0}^{\tau _{0}}d\tau \tau
V_{sc}(\tau )\right\} .  \nonumber
\end{eqnarray}
To discuss the result, we for a moment assume a simple $\tau
$-dependence of
$V_{ij}(\tau )$
\begin{equation}
V_{ij}(\tau )=V_{ij}(0)(1-{\frac{\tau }{\tau _{0}}}),  \label{eq:a3}
\end{equation}
which corresponds to the $r$-dependence used in Eq.~(\ref{eq:ma6a}). We
note, however, that this form is too simple to describe the behavior of
the
more realistic potential in Eq.~(\ref{eq:m4}). Inserting
Eqs.~(\ref{eq:a1a})
and (\ref{eq:a3}) in Eq.~(\ref{eq:a1b}) gives
\begin{eqnarray}\label{eq:a4}
\Delta _{2} &\equiv |&a_{2v}(\tau
_{0})|^{2}-|a_{2v}(0)|^{2}={\frac{1}{4}}%
V_{sc}(0)\sin (2\varphi )  \nonumber   \\
&&\times \left[ {\frac{1}{4}}\sin (2\theta )V_{sc}(0)+{\frac{1}{3}}\sin
(2\varphi +2\theta )\delta E\right] \tau _{0}^{2}.
\end{eqnarray}

We now extend this treatment to intermediate values of $v$ where the
adiabatic to sudden transition takes place. Using the expressions
Eqs.~(\ref
{eq:s2}), (\ref{eq:s5}) and (\ref{eq:s7}) to relate $\delta E$ and $%
U=-V_{sc}(0)$ we obtain
\begin{equation}
\frac{r(\omega )}{r_{00}}-1=\frac{\Delta _{2}}{\sin ^{2}(2\varphi )\cos
^{2}(2\varphi )}=-{\frac{\sin (2\varphi )}{\sin (2\theta
)}}{\frac{(\delta
E)^{2}\tau _{0}^{2}}{12}}.  \label{eq:a5}
\end{equation}
{\it Within the semi-classical theory} the condition for the sudden
approximation is then
\begin{equation}
{\frac{v}{R_{0}}}={\frac{1}{\tau _{0}}}\gg \delta E\sqrt{\frac{\sin
(2\varphi )}{12\sin (2\theta )}}.  \label{eq:a6}
\end{equation}

We are now in a position to discuss the approach to the sudden limit.
Within
a semi-classical framework it seems clear that we have to require that
the
hole potential is fully switched on after a ``short'' time $\tau _{0}={%
v/R_{0}}$, i.e., that the emitted electron leaves the range of the
scattering potential after a short time. The question is, however, what
we
mean by ``short''. From Eq.~(\ref{eq:a4}) it follows that the time-scale
is
set by both the inverse of $\delta E$ and the inverse of $V_{sc}(0)$. In
Eq.~(\ref{eq:a6}) we have used the relation between $\delta E$ and
$V_{sc}(0)
$ to remove $V_{sc}(0)$ from Eq.~(\ref{eq:a6}).

 From Eq.~(\ref{eq:a6}) we obtain the condition for the sudden
approximation
within the SC theory
\begin{equation}  \label{eq:a6a}
\tilde{\omega}={\frac{1}{2}}k^2\gg (\delta E)^2 R_0^2={\frac{(\delta
E)^2}{2%
\tilde E}}\sim {\frac{V_{sc}(0)^2 }{\tilde E}}.
\end{equation}
Thus, according to the semi-classical theory, the sudden approximation
requires that $\epsilon_k\gg(\delta E)^2/\tilde E$.
Comparison with the full quantum mechanical calculations in Fig. 
\ref{fig:strength} shows that this criterion is not appropriate 
for the range of parameters considered here. The reason is that we 
have considered a parameter range where the semi-classical theory
is not very accurate.

It is interesting that the SC theory correctly predicts that the 
weight of the satellite goes to zero at threshold. Nevertheless, 
the SC theory does not give the correct physics at the threshold. 
In the full quantum mechanical calculation the weight of the 
satellite goes to zero due to the effects of the dipole matrix 
element, which becomes very small at small photoelectron energies. 
This effect is not included in the SC theory. In the semi-classical 
treatment, the small weight of the satellite is due to the fact 
that the scattering potential between the outgoing slow electron and
the excitation means that the hole potential is only switched on slowly.
In the quantum mechanical treatment, on the other hand, the scattering
potential leads to an enhancement of the relative weight of the
satellite close to the threshold for the shake-down case.

\section{Discussion}

We have studied the photoemission spectrum of a simple model with a
localized charge transfer excitation. We have obtained exact numerical
results for the spectrum as a function of the photon energy $\omega$ and
in particular focussed on the ratio $r(\omega)$ between the weights of 
the satellite and the main peak. These calculations are compared with
perturbational and semi-classical treatments. The results have been
analyzed using the latter two approaches.

An important effect in the ratio $r(\omega)$ is due to the energy
dependence of the dipole matrix elements and a factor $1/(\partial
\epsilon_k/\partial k)\sim 1/k$ in the expression for the spectrum. 
This leads to a suppression
of the satellite close to the threshold, but can lead to an overshoot
further away from the threshold. This effect was discussed in Sec. \ref
{sec:sudden} and is described by $r_0(\omega)$. If the interaction
between the emitted electron and the target is weak, this effect 
dominates. It is determined by the excitation energy $\delta E$ and 
the relevant energy scale $\tilde E_d$ of the dipole matrix element. 
If $\delta E/\tilde E_d$ is small, $ r_0$ reaches its limiting value 
from below, while there is an overshoot if $\delta E/\tilde E_d\gg 1$. 
In both cases $r_0$ reaches its limiting value for
a kinetic energy of the order a few times $\delta E$.

To study the effects of the scattering potential between the emitted
electron and the target we have focussed on the ratio
$r(\omega)/r_0(\omega)$. This quantity shows an overshoot for small 
values of $\omega$ in the ``shake-down'' situation studied here. 
Depending on the parameters there may
be an undershoot for larger energies, which extends over a large energy
range. This undershoot is, however, fairly small for the cases
considered here. The sudden approximation is then valid to a 
reasonable accuracy when
the overshoot has become small. We show that this happens on the energy
scale $\tilde E=1/(2\tilde R^2)$, where $\tilde R$ is a typical length
scale of the scattering potential.

One of the main results of this paper is that for a coupling
to localized excitations, the adiabatic to sudden transition 
takes place at quite small kinetic energies of the photoelectron.
This is in contrast to the large kinetic energies needed for the 
case of coupling to plasmons. In the plasmon case, the kinetic 
energy is typically so large that the semi-classical treatment is 
a very good approximation. The adiabatic to sudden transition is 
then expected to happen on the energy scale 
$(\omega_q\lambda)^2$,\cite{Inglesfield} where $\omega_q$ and 
$\lambda$ are the plasmon frequency and wavelength, respectively.
Since the long wavelength plasmons dominate the transition, this
happens at very large energies. For a localized excitation, the 
relevant length scale of the scattering potential is smaller, and 
the transition is expected to take place at a smaller energy scale. 
Actually, the transition takes place at such a small energy that 
the semi-classical theory is usually not valid any more. It is 
interesting that the semi-classical theory therefore predicts the 
opposite dependence on the range $\tilde R$ of the scattering
potential, namely as $\tilde R^2$ instead of $\tilde E =1/(2\tilde
R^2)$.

For the ''shake-down'' scenario considered here (the two outer levels
cross as the hole is created), we find constructive interference 
(increase of $r(\omega )/r_{0}(\omega )$) between the intrinsic and 
extrinsic processes at
low photoelectron energies. This is in contrast to the destructive
interference found in the plasmon case and to the reduction of $r(\omega
)/r_{0}(\omega )$ found here for the ''shake-up'' case (no level
crossing).

\section*{acknowledgement}

This work has been supported by the Max-Planck Forschungspreis. 
One of us (LH) carried out part of his contribution to this work 
at the Max-Planck Institute for Festk\"{o}rperforschung.

\begin{table}[t]
\caption{The model Hamiltonian Eq.(\ref{eq:1}) can describe various
charge transfer systems. The table indicates the meaning of the 
states $a$ and $b$ for different cases.}
\label{general-CT}
\begin{tabular}{ccc}
& a & b \\ \hline
&  &  \\[-2mm]
transition metal compounds & $3d$ state & ligand state \\[2mm]
CO on surface & $2\pi^{\ast}$ state & bulk(surface) state \\[2mm]
Ce compounds & $4f$ state & $5d$ state
\end{tabular}
\end{table}

\begin{table}[tbp]
\caption{Used parameters for copper-dihalide compounds.}
\label{used-parameter}
\begin{tabular}{ccccccc}
& $U$ (eV) & $t$ (eV) & $\delta E$ (eV) & $R_0$ (a.u.) & $\varepsilon$ &
$%
r(\omega\rightarrow\infty)$ \\ \hline
&  &  &  &  &  &  \\[-2mm]
CuBr$_2$ & 12.33 & 2.02 & 7.37 & 5.01 & 1.71 & 2.33 \\[2mm]
CuCl$_2$ & 10.58 & 1.84 & 6.45 & 4.71 & 1.96 & 2.07 \\[2mm]
CuF$_2$ & 8.62 & 1.63 & 5.41 & 3.86 & 2.26 & 1.75
\end{tabular}
\end{table}

\end{multicols}


\begin{references}
\bibitem{Kaupp}  M. Kaupp, private communications; for example, in the
atomic configuration of the Cu-Cl octahedral cluster in CuCl$_{2}$, $%
l_{c}=5.59$ a.u., $l_{a}=l_{b}=4.27$ a.u., and the average distance of
them will $R_{0}=4.71$ a.u. (see the inset of Fig.\ref{fig:potential}a).
\end{references}
\end{document}